\documentclass[a4paper,11pt]{article}
\pdfoutput=1 

\usepackage{jheppub} 
                     
\usepackage[T1]{fontenc} 
\usepackage{xcolor}
\usepackage{subfigure}
\usepackage{braket}
\usepackage{bbm}
\usepackage{hyperref}

\newcommand{\be}{\begin{equation}}
\newcommand{\ee}{\end{equation}}
\newcommand{\bea}{\begin{eqnarray}}
\newcommand{\eea}{\end{eqnarray}}

\begin{document}

\title{\boldmath Symmetry-resolved modular correlation functions in free fermionic theories
}

\vspace{.5cm}

\author{Giuseppe Di Giulio and Johanna Erdmenger}
\affiliation{Institute for Theoretical Physics and Astrophysics and W\"urzburg-Dresden Cluster of Excellence ct.qmat, Julius-Maximilians-Universit\"at W\"urzburg, Am Hubland, 97074 W\"{u}rzburg, Germany.}

\emailAdd{giuseppe.giulio@physik.uni-wuerzburg.de}

\vspace{.5cm}

\abstract{As a new ingredient for analyzing the fine structure of entanglement, we study the symmetry resolution of the modular flow  of $U(1)$-invariant operators in theories endowed with a global $U(1)$ symmetry.
We provide a consistent definition of symmetry-resolved modular flow that is defined for a local algebra of operators associated to a sector with fixed charge.
We also discuss the symmetry-resolved modular correlation functions and show that they satisfy the KMS condition in each symmetry sector.
Our analysis relies on the factorization of the Hilbert space associated to spatial subsystems.
We provide a toolkit for computing the symmetry-resolved modular correlation function of the charge density operator in free fermionic theories. As an application, we compute this correlation function for a $1+1$-dimensional free massless Dirac field theory and find that it is independent of the charge sector at leading order in the ultraviolet cutoff expansion. This feature can be regarded as a charge equipartition of the modular correlation function.
Although obtained for free fermions, these results may be of potential interest for bulk reconstruction in AdS/CFT.
}

\maketitle

\newpage

\section{Introduction}
\label{sec:Intro}

In the last decades a remarkable effort has been made towards understanding entanglement in quantum field theories (QFTs). Many insights were provided tackling the problem from a formal perspective, describing field theories in terms of local algebras of bounded operators \cite{Haagbook,Petzbook} (see also \cite{Witten:2018zxz} for a recent review). In this framework, known as algebraic quantum field theory (AQFT), several important concepts were introduced, a particularly insightful one being the {\it modular flow}. Roughly speaking, the modular flow is a generalized time evolution induced by a reduced density matrix $\rho_V$ of a given spatial region $V$ on any operator $A$ localized in the same region, namely $\sigma_t(A)\equiv \rho_V^{\mathrm{i}t}A\rho_V^{-\mathrm{i}t}$. 
An important property is that the correlation functions of operators evolved along this flow, called {\it modular correlation functions}, satisfy a periodicity condition for imaginary values of the parameter $t$, which is known as Kubo-Martin-Schwinger (KMS) condition.
Motivated by the progress in understanding entanglement achieved through its symmetry resolution, this work is devoted to the study of the interplay between symmetries and modular flow. In \cite{Erdmenger:2020nop} a resolvent method was proposed that allows to obtain the modular flow for free fermionic theories without employing the explicit expression for the modular Hamiltonian. In the present paper, we adapt this approach to the computation of the modular flow of the charge density operator. This is done in view of studying the modular flow in different symmetry sectors of the theory.

The modular flow and the modular correlation functions have applications in the context of modular theory \cite{Lashkari:2018nsl,Lashkari:2019ixo}, entropy and energy inequalities \cite{Casini:2008cr,Blanco:2013lea,Faulkner:2016mzt,Balakrishnan:2017bjg,Ceyhan:2018zfg} and the AdS/CFT correspondence \cite{Maldacena:1997re,Witten:1998qj,Gubser:1998bc}. The modular flow was  an important ingredient in unraveling the relation between entanglement and spacetime geometry in holography, in particular in the context of the Ryu-Takayanagi formula \cite{Ryu:2006bv} and the developments it triggered.
 Indeed, as elucidated in \cite{Jafferis:2014lza,Jafferis:2015del}, a holographic dual of the modular flow in a given region of the boundary conformal field theory (CFT) can be identified, helping in understanding the deep relation between a generic boundary subregion and the corresponding entanglement wedge in the bulk.
Moreover, the modular flow has been used to shed light on aspects such as the bulk locality in AdS/CFT and the bulk reconstruction program \cite{Hamilton:2006az,Kabat:2011rz,Faulkner:2017vdd,Faulkner:2018faa,DeBoer:2019kdj}, while the modular correlation functions of CFT operators can be exploited as boundary probes of quantum extremal surfaces in the bulk
\cite{Chandrasekaran:2021tkb,Chandrasekaran:2022qmq}.

The explicit expression of the modular flow of a given spatial subregion and the corresponding modular correlation functions are known only in a few cases. The most important ones concern theories in their vacuum states restricted to the Rindler wedge for a generic relativistic QFT \cite{Bisognano:1975ih,Bisognano:1976za} and a spherical subregion in a CFT \cite{Hislop:1981uh,Casini:2011kv,Wong:2013gua,Cardy:2016fqc}. Beyond these cases, the results strongly depend on the details of the theory. A simple example for which results were obtained also for more complicated bipartitions is the one of free fermionic theories.
The entanglement properties of the $1+1$-dimensional free massless Dirac theory bipartite into a region made up of $p$ disjoint intervals and its complement have been first studied in \cite{Casini:2005rm}. In \cite{Casini:2009vk}, the modular Hamiltonian, i.e. the logarithm of the reduced density matrix, of this bipartition was worked out and the corresponding modular flow and modular correlation functions were computed. These findings have then been generalized to other bipartitions of free fermionic theories \cite{Klich:2015ina,Arias:2017dda,Arias:2018tmw,Fries:2019ozf,Blanco:2019xwi,Erdmenger:2020nop,Mintchev:2020uom,Rottoli:2022plr,Chen:2022nwf}.

Another evergreen subject of research concerns the role played by 
symmetries in understanding physical systems and phenomena. Recently, various experimental \cite{Lukin19,Azses:2020tdz,Neven:2021igr,Vitale:2021lds} and theoretical \cite{LaFlorencie2014,Goldstein:2017bua,Xavier:2018kqb} results have motivated an intense investigation of the relation between symmetries and entanglement.
The main tools developed for this purpose are the {\it symmetry-resolved entanglement entropies}, which can be roughly thought as the entanglement and Rényi entropies associated to the different charged sectors of theories endowed with a given global symmetry.
These quantities have been computed in several cases in QFTs \cite{Goldstein:2017bua,Xavier:2018kqb,Murciano:2020vgh,Horvath:2020vzs,Capizzi:2020jed,Bonsignori:2020laa,Estienne:2020txv,Calabrese:2021wvi,Milekhin:2021lmq,Ma:2021zgf,Capizzi:2021kys,Horvath:2021fks,Horvath:2021rjd,Capizzi:2022jpx,Capizzi:2022nel,Ares:2022gjb,Ghasemi:2022jxg,DiGiulio:2022jjd,Foligno:2022ltu,Capizzi:2023bpr,Fossati:2023zyz,Zhao:2020qmn,Weisenberger:2021eby,Zhao:2022wnp} and, in all the analyzed instances, an intriguing feature has been observed: the symmetry-resolved entanglement entropies are independent of the charge sector at leading order in the ultraviolet (UV) cutoff expansion. This feature was dubbed entanglement equipartition \cite{Xavier:2018kqb}. 
The entanglement entropies and their averages in statistical ensembles \cite{Murciano:2022lsw} are not the only quantities whose symmetry resolution was investigated; other examples include negativity \cite{Cornfeld:2018wbg,Murciano:2021djk,Chen:2021nma,Chen:2022gyy,Parez:2022xur}, relative entropies and distances \cite{Chen:2021pls,Capizzi:2021zga} and operator entanglement \cite{Wellnitz:2022cuf,Rath:2022qif}.

In this work we consider theories with a $U(1)$ global symmetry such that, for each subregion $V$, we can define the $U(1)$ charge $Q_V$ restricted to $V$.
We decompose a given local algebra associated to $V$ into the various sectors at fixed charge $Q_V$ and, for each fixed-charge algebra, we formulate the Tomita-Takesaki theory. This allows to define a full-fledged {\it symmetry-resolved modular flow} and the corresponding {\it symmetry-resolved modular correlation functions}. Remarkably, this can only be achieved by considering $U(1)$ invariant operators and therefore not all the possible modular flows and modular correlation functions can be resolved into charge sectors.
Consistently with the idea of symmetry resolution, by properly summing over all the possible charge sectors, we recover the unresolved modular flow and the corresponding modular correlation functions.
Our analysis holds for a general theory as long as the Hilbert space can be factorized into the contributions from the degrees of freedom in a spatial subsystem and its complement. The factorization of the Hilbert space occurs for quantum theories described in terms of local finite-dimensional algebras. This factorization formally provides a tool for accessing well-defined quantities such as the modular flow also for free QFTs \cite{Araki:1971id,Hollands:2019hje}.
To the best of our knowledge, the symmetry resolution of the modular flow and the modular correlation functions has not been investigated before.

For the sake of concreteness, we focus on fermionic Gaussian states in generic dimension, an important class of states which includes also the ground states of free fermionic theories. We adapt the computation method of \cite{Erdmenger:2020nop} for the modular flow and the modular correlation function to the conserved $U(1)$ charge density, retrieving the known expression for this flow and correlator \cite{Hollands:2019hje, Mintchev:2020jhc}.
In \cite{Erdmenger:2020nop},  the analytic structure, i.e. the poles, of the modular correlation function  has been related to the locality properties of the modular flow for an elementary fermionic field. We confirm that this relation also holds for the charge density operator.
Given that the modular correlation function of the charge density operator turns out to be the square of that of an elementary field, its analytic structure  becomes richer.

Moreover, we develop a strategy for computing the symmetry resolution of the modular correlation function of the charge density operator. Inspired by a known paradigm in the context of symmetry-resolved entanglement \cite{Goldstein:2017bua}, we find that the computation boils down to calculate the Fourier transform of a specific correlation function.
We carry out this computation explicitly for the $1+1$-dimensional free massless Dirac field theory on an infinite line, bipartite into a region made up of $p$ intervals and its complement. The symmetry-resolved modular correlation function is achieved by exploiting bosonization techniques. An intriguing finding is that the modular correlation function in each charge sector does not depend on the value of the charge at leading order in the UV cutoff expansion. We can think of this property as an {\it equipartition of the modular correlation function of the charge density}. The dependence on the charge sector is then observed in subleading terms, which vanish as the UV cutoff is sent to zero.

These results are of potential interest for the AdS/CFT correspondence, since they provide insights on the charge decomposition of the density operator, which is an invariant composite operator. This may provide further useful ingredients for bulk reconstruction, in addition to for instance \cite{Foit:2019nsr,Johnson:2022cbe}. In particular, it may be envisioned to generalize the results presented here for the charge density to higher-spin conserved currents \cite{Gaberdiel:2010pz, Bekaert:2014cea}.  Moreover, as we comment further in the conclusion and outlook section below, our results for the modular flow may also be of interest for the recently proposed application of von Neumann algebras to the AdS/CFT correspondence \cite{Jefferson:2018ksk,Leutheusser:2021qhd,Witten:2021unn,Chandrasekaran:2021tkb}.

The paper is organized as follows. In Sec.\,\ref{sec:AlgebraicModularTheory}, we first briefly review concepts of algebraic quantum field theory that will be useful in the forthcoming discussion. We then perform an analysis of the modular theory in different $U(1)$ charge sectors, which leads to the definition of symmetry-resolved modular flow and symmetry-resolved modular correlation functions.
A strategy for computing the latter is discussed in Sec.\,\ref{sec:FreeFermions} in the context of fermionic Gaussian states and an explicit calculation for the $1+1$-dimensional free massless Dirac theory is reported in Sec.\,\ref{sec:ChiralFF2d}.
Finally, in Sec.\,\ref{sec:Conclusion} we summarize the main findings and we comment on possible extensions of this work. We also include four appendices in which computational details and additional results are reported.

\section{Modular flow and charge decomposition}
\label{sec:AlgebraicModularTheory}

In this section we review the aspects of algebraic quantum field theory that will be useful throughout the manuscript, as the Tomita-Takesaki modular theory and the properties of twist operators in $U(1)$-symmetric theories. Based on these concepts, we restrict our analysis to theories endowed with a global $U(1)$ symmetry.  For theories defined in terms of finite-dimensional local algebras of operators, we introduce a definition of symmetry-resolved modular flow and symmetry-resolved modular correlation functions of $U(1)$-invariant operators.

\subsection{Tomita-Takesaki theory}
\label{subsec:TTtheory}

In the algebraic description of a quantum theory a key role is played by algebras of operators.
In an algebraic quantum field theory, we associate von Neumann (vN) algebras of operators to bounded regions of space. The elements of a vN algebra can be represented as a subset of the set $\mathcal{B}(\mathcal{H})$ of all the bounded operators acting on the Hilbert space $\mathcal{H}$ of the QFT. More precisely, an algebraic quantum field theory is defined by a {\it net of local algebras}, which associate a vN algebra $\mathcal{A}(V)\subset \mathcal{B}(\mathcal{H})$ to each causally complete spacetime region $V$. This net of algebras must satisfy a list of axioms; we refer the interest reader to \cite{Haag:1967sg} for the complete list and discussion of these axioms.

Consider a spatial region $V$ and the associated algebra $\mathcal{A}(V)\subset \mathcal{B}(\mathcal{H})$. In AQFT the states are defined as linear applications from the operator algebra to the complex numbers and can be represented as vectors in a Hilbert space. We consider the state $|\Omega\rangle\in\mathcal{H}$ and we assume it is {\it cyclic and separating} for $\mathcal{A}(V)$. A state vector $|\Omega\rangle$ is said cyclic for $\mathcal{A}(V)$ if $\mathcal{A}(V)|\Omega\rangle$ is dense in $\mathcal{H}$, while it is separating for $\mathcal{A}(V)$ if $\nexists A \in \mathcal{A}(V)$ such that $A|\Omega\rangle=0$. 
%
It is possible to show that, if $V$ is a non-empty open region with a non-empty causal complement, the vacuum state of a relativistic QFT is always cyclic and separating \cite{Arakibook} and therefore the forthcoming analysis applies to this state.
Given the cyclic and separating state $|\Omega\rangle$, there exists a unique antilinear {\it modular involution} $S_\Omega$ such that \cite{Brattelibook}
\be
\label{eq:Modular relation general}
S_\Omega A |\Omega\rangle= A^\dagger|\Omega\rangle,
\qquad
\forall A\in \mathcal{A}(V).
\ee
The polar decomposition
\be
\label{eq:polardecompositiongeneral}
S_\Omega=J_\Omega \Delta^{1/2}_\Omega
\ee
allows to introduce the antiunitary modular conjugation $J_\Omega$ and the positive, self-adjoint {\it modular operator} $\Delta_\Omega$.
 As we will discuss later in this section, when $\mathcal{A}(V)$ is a finite dimensional vN algebra, the modular operator is related to the reduced density matrix of the subregion $V$. 
A fundamental result of the modular theory is the Tomita-Takesaki theorem, which says that \cite{Haagbook} 
\be
\label{eq:TTtheorem}
\Delta_\Omega^{\mathrm{i}t}\mathcal{A}(V)\Delta_\Omega^{-\mathrm{i}t}=\mathcal{A}(V),\qquad
\forall t \in \mathbb{R}\,.
\ee
For any $t\in \mathbb{R}$ and $ A\in\mathcal{A}(V)$, $\Delta_\Omega^{\mathrm{i}t}A\Delta_\Omega^{-\mathrm{i}t}$ is called {\it modular flow}. The theorem (\ref{eq:TTtheorem}) can be rephrased saying that the modular flow preserves the algebra $\mathcal{A}(V)$.

The modular flow allows to construct the so-called {\it modular correlation function}
\begin{equation}
  \label{eq:modcorrfunc_AQFT}
    G_{\textrm{\tiny mod}}(A,B;t)
    \equiv
\langle\Omega|B\Delta_\Omega^{\mathrm{i}t}A\Delta_\Omega^{-\mathrm{i}t}|\Omega\rangle,
    \qquad
    A,B\in\mathcal{A}(V),
    \quad
    t\in\mathbb{R}\,.
\end{equation}
This quantity as function of $t$ is analytic in the strip 
$-1<\textrm{Im}(t)<0$ and can be analytically continued to $0<\textrm{Im}(t)<1$ by defining
\begin{equation}
\label{eq:KMS_AQFT}
 G_{\textrm{\tiny mod}}(A,B;t+\mathrm{i}) 
 =
\langle\Omega|\Delta_\Omega^{\mathrm{i}t}A\Delta_\Omega^{-\mathrm{i}t}B|\Omega\rangle\,.
\end{equation}
This continuation comes from the {\it Kubo-Martin-Schwinger (KMS) condition} \cite{Haagbook}, which can be alternatively understood as the fact that the modular flow behaves like a time evolution with respect to a Hamiltonian $-K=\ln \Delta_\Omega $ in a thermal state with temperature $-1$. The operator $K$ is called {\it modular Hamiltonian}.

Let us specialize the above discussion to finite-dimensional algebras, namely when the operators in $\mathcal{A}(V)$ can be represented as matrices.
In this case, we can consider the Hilbert space $\mathcal{H}$ of our theory to be factorized as $\mathcal{H}=\mathcal{H}_V\otimes \mathcal{H}_{V'}$, where the local algebra $\mathcal{A}(V)$ of operators 
located in the region $V$ acts non-trivially only on $\mathcal{H}_V$, while the algebra $\mathcal{A}(V')$ of operators in the complementary region $V'$ only on $\mathcal{H}_{V'}$. In other words, $A\in\mathcal{A}(V)$ is represented on $\mathcal{H}$ as $A\otimes\boldsymbol{1}_{V'}$, while $A'\in\mathcal{A}(V')$ as $\boldsymbol{1}_{V}\otimes A'$. Notice that the algebras $\mathcal{A}(V)$ and $\mathcal{A}(V')$ are each other's commutants.
 Following the steps of the modular theory for a finite dimensional algebra, we have \cite{Haagbook}
\be
\label{eq:modopRDM0}
\Delta_\Omega=
\rho_{V}\otimes\rho_{V'}^{-1}
,
\ee
and 
\be
\label{eq:modflowRDM}
\Delta_\Omega^{\mathrm{i}t}A\Delta_\Omega^{-\mathrm{i}t}
=\rho_{V}^{\mathrm{i}t}A\rho_{V}^{-\mathrm{i}t}\otimes \boldsymbol{1}_{V'}
\equiv
\sigma_t(A)\otimes \boldsymbol{1}_{V'}
\,,
\ee
which, as anticipated, provides a connection between the modular operator and the reduced density matrix $\rho_V$. In the finite-dimensional setting, $\rho_V$ can be obtained by computing the partial trace over $\mathcal{H}_{V'}$ of the density matrix $|\Omega\rangle\langle\Omega|$ of the entire system.
Notice that $\rho_V$ obviously depends on the state $|\Omega\rangle$, but we omit this explicit dependence to lighten the notation.
It is worth stressing that the reduced density matrices do not exist as operators in algebraic quantum field theory. In the rest of this work we restrict our analysis to finite dimensional algebras, for which (\ref{eq:modopRDM0}) and (\ref{eq:modflowRDM}) hold and the reduced density matrices are well-defined. This is by no mean unrelated to QFT.
Indeed, it is possible to prove that infinite-dimensional algebras, as the ones associated to quantum field theories, can be obtained as limits of successions of matrix algebras \cite{Longo78}.
Moreover, it has been shown that in free QFTs it is often convenient to formally introduce reduced density matrices, that, although are not well-defined, can be used as tools to achieve other meaningful results as modular flow and modular correlation functions \cite{Araki:1971id,Hollands:2019hje}. In Sec.\,\ref{sec:FreeFermions}
 and Sec.\,\ref{sec:ChiralFF2d}, we apply this idea in the context of free fermionic field theories.

\subsection{Twist operators and projectors in \texorpdfstring{$U(1)$}{U(1)} charge sectors}
\label{subsec:twistchargesector}

Consider now a net of local algebras that we refer to as {\it field algebra}, 
which fulfils all the axioms for defining a QFT.
Consider then the compact abelian group $U(1)$.
We denote by $g_\alpha$, with $\alpha\in[-\pi,\pi)$, a unitary representation of the elements of $U(1)$ on the Hilbert space $\mathcal{H}$.
We assume that all the elements of $U(1)$ commute with all the elements of the Poincaré group. Thus, since the local algebras must satisfy Poincaré covariance, the elements of $U(1)$ do not affect the spacetime points,
meaning that the group $U(1)$ implements internal symmetries.
The net of {\it local observable algebras} $\mathcal{A}$ is defined as the algebra of the operators in the field algebra that commutes with every element of $U(1)$. In other words, $\mathcal{A}$ is the $U(1)$-invariant part of the field algebra.
 We denote $\mathcal{A}(V)\subset \mathcal{A}$ the local algebras in this net.
Remarkably, if the field algebra defines a genuine QFT, also $\mathcal{A}$ does \cite{Doplicher:1969tk,Doplicher:1969kp}.
An important consequence of the $U(1)$ symmetry of the theory is that the Hilbert space can be decomposed as
\be
\label{eq:decHS}
\mathcal{H}=\bigoplus_{q\in\mathbb{Z}}\mathcal{H}_{q}\,,
\ee
where each term in the direct sum is associated to a different eigenvalue of the charge operator $Q$, i.e. the generator of the $U(1)$ symmetry.
We assume now that the cyclic and separating vector $|\Omega\rangle$ is an eigenstate of $Q$ with eigenvalue $\bar{q}$. This means that, given $g_\alpha\in U(1)$, $g_\alpha|\Omega\rangle=e^{\mathrm{i}\alpha\bar{q}}|\Omega\rangle$, $\forall \alpha\in[-\pi,\pi)$. 
From the viewpoint of the decomposition (\ref{eq:decHS}), this means that $|\Omega\rangle\in\mathcal{H}_{\bar{q}}$. Let us stress that the vacuum of a $U(1)$-invariant theory belongs to the subspace $\mathcal{H}_0$. 

With the purpose of realizing a given symmetry in a subregion $V$ of the spacetime, the notion of {\it twist operators} has been introduced \cite{Doplicher:1972kr}. In general, the twist operators for the $U(1)$ symmetry are defined in such a way that they are equal to $g_\alpha$ in the region $V$ and act as the identity in an open region contained in the complement of $V$.
In the finite-dimensional setting we consider in this manuscript, the Hilbert space $\mathcal{H}_{\bar{q}}$, as discussed for $\mathcal{H}$ in the previous section, factorizes as $\mathcal{H}_{\bar{q},V}\otimes \mathcal{H}_{\bar{q},V'}$ and the restriction $Q_V\otimes\boldsymbol{1}_{V'}\in\mathcal{A}(V)$ of $Q$ to the subregion $V$ is well-defined. Thus, the twist operators can be written as
\begin{equation}
 \tau_{V}(\alpha)\otimes\boldsymbol{1}_{V'}\in \mathcal{A}(V)\,,
\qquad\qquad
   \tau_{V}(\alpha)\equiv e^{\mathrm{i}\alpha Q_V}\,,
   \qquad\qquad
\alpha\in [-\pi,\pi)\,.
\label{eq:twistoperatordef}
\end{equation}
This definition can be straightforwardly modified for the twist operators associated to the complementary region $V'$, which belong to the algebra $\mathcal{A}(V')$. 
\\
The twist operators realize a unitary representation of $U(1)$ and
 can be decomposed into irreducible representations as
\begin{equation}
\label{eq:decompositiontwist}
\tau_{V}(\alpha)=\sum_{q\in\mathbb{Z}}e^{\mathrm{i}\alpha q}\Pi_V(q)\,,
\end{equation}
where $\Pi_V(q)$ is an operator projecting onto the eigenspace of $Q_V$ with eigenvalue $q$. Meant as operator in $\mathcal{H}_{\bar{q},V}$, $\Pi_V(q)$ is a projector onto a subspace that we call $\mathcal{H}^{(q)}_{V}$. The same decomposition (\ref{eq:decompositiontwist}) can be written down also for the twist operator $\tau_{V'}(\alpha)$ associated to the complement of $V$ and allows to introduce the projector $\Pi_{V'}(q)$ onto $\mathcal{H}^{(q)}_{V'}$. Notice that both $\mathcal{H}^{(q)}_{V} $ and $\mathcal{H}^{(q)}_{V'} $ are subspaces of $ \mathcal{H}_{\bar{q},V}$ and $ \mathcal{H}_{\bar{q},V'} $ respectively and depend on $\bar{q}$. We do not report explicitly this dependence to avoid clutter. Taking into account that the restriction to $\mathcal{H}_{\bar{q}}$ fixes the total charge of the system to be $\bar{q}$, we obtain the decomposition
\be
\label{eq:Hq_dec}
\mathcal{H}_{\bar{q}}=\bigoplus_{q\in\mathbb{Z}}\mathcal{H}_{V}^{(q)}\otimes \mathcal{H}_{V'}^{(\bar{q}-q)}
\,,
\ee
where the various subspaces can be accessed using the projectors $\Pi_V$ and $\Pi_{V'}$ introduced above.
A consequence of \eqref{eq:Hq_dec} is that, given a state $|\Omega\rangle\in\mathcal{H}_{\bar{q}}$, we have
\be
\label{eq:projectorsonGS}
\Pi_{V}(q)\otimes\Pi_{V'}(\bar{q}-q)|\Omega\rangle=\Pi_{V}(q)\otimes\boldsymbol{1}_{V'}|\Omega\rangle\,,
\ee
which can be physically understood by noticing that, since the total charge in the state $|\Omega\rangle$ is $\bar{q}$, once we project onto the subspace with charge $q$ in the subsystem $V$, the charge in $V'$ is constrained to be $\bar{q}-q$. 
%
For later convenience, it is useful to notice that the infinite sum in (\ref{eq:decompositiontwist}) can be inverted and the projectors read
\begin{equation}
\label{eq:FourierTransform}
\Pi_V(q)=\int_{-\pi}^\pi\frac{\mathrm{d}\alpha}{2\pi}e^{-\mathrm{i}\alpha q}\tau_V(\alpha)\,,
\end{equation}
and similarly for the projectors in the algebra $\mathcal{A}(V')$.

The projector in (\ref{eq:FourierTransform}) can be exploited for achieving a charge decomposition of the algebra $\mathcal{A}(V)$. Indeed, 
the elements of $\mathcal{A}(V)$ commute with the elements of $U(1)$ and the latter are represented in the region $V$ as $\tau_V(\alpha)\otimes\boldsymbol{1}_{V'}$ with 
$\alpha\in [-\pi,\pi)$, meaning that, if $A\otimes \boldsymbol{1}_{V'}\in\mathcal{A}(V)$, $[A,\tau_V(\alpha)]=0$ and therefore $[A,\Pi_V(q)]=0$. The last commutator allows to write $A$ in the following decomposed form
\begin{equation}
\label{eq:Adecomposition}
    A=\sum_{q\in \mathbb{Z}} \Pi_V(q) A \Pi_V(q)\equiv \sum_{q\in \mathbb{Z}} A_q\,,
\end{equation}
which implies that the algebra $\mathcal{A}(V)$ can be decomposed as the direct sum
\begin{equation}
\label{eq:decompositionalgebra}
    \mathcal{A}(V)=\bigoplus_{q\in\mathbb{Z}} \mathcal{A}_q(V).
\end{equation}
From the definition of $A_q$ in (\ref{eq:Adecomposition}), we can see that 
$\mathcal{A}_q(V)$ acts non-trivially only on the Hilbert space $\mathcal{H}_{V}^{(q)}$ defined in (\ref{eq:Hq_dec}). In the following, we represent $\mathcal{A}_q(V)$ as an algebra of bounded operators acting on $\mathcal{H}^{(q)}_{V}\otimes \mathcal{H}^{(\bar{q}-q)}_{V'}$.
It is worth commenting that, given $|\Omega\rangle\in\mathcal{H}_{\bar{q}}$, the corresponding reduced density matrices $\rho_V$ and $\rho_{V'}$
introduced in (\ref{eq:modopRDM0}) commute with $Q_{V}$ and $Q_{V'}$ respectively and therefore also with $\Pi_{V}(q)$ and $\Pi_{V'}(q)$, for any $q$. 
In other words, $\rho_V\otimes \boldsymbol{1}_{V'}\in\mathcal{A}(V)$ and $\boldsymbol{1}_{V}\otimes\rho_{V'}\in\mathcal{A}(V')$. 
This fact will be crucial in the forthcoming sections.

\subsection{Symmetry resolution of modular flow and modular correlation functions}
\label{subsec:AQFT_srmodflow}

The goal of this section is to achieve a symmetry resolution of the modular flow defined in (\ref{eq:TTtheorem}) and of the modular correlation function in (\ref{eq:modcorrfunc_AQFT}).
We assume the validity of the modular relation (\ref{eq:Modular relation general}) with $A\otimes\boldsymbol{1}_{V'}\in\mathcal{A}(V)$ and $|\Omega\rangle\in \mathcal{H}_{\bar{q}}$ and we look for a symmetry-resolved modular relation of the form
\begin{equation}
\label{eq:modularrelationfixedq}
    S_{\Omega,q} \left(A_q\otimes \boldsymbol{1}_{V'}\right) |\Omega_q\rangle= A_q^\dagger\otimes \boldsymbol{1}_{V'}|\Omega_q\rangle\,,
\end{equation}
where $A_q\otimes \boldsymbol{1}_{V'}\in\mathcal{A}_q(V)$ and $|\Omega_q\rangle\in \mathcal{H}_{V}^{(q)}\otimes\mathcal{H}_{V'}^{(\bar{q}-q)}$.
If (\ref{eq:modularrelationfixedq}) holds, $S_{\Omega,q}$ allows to define a symmetry-resolved modular operator associated to the algebra $\mathcal{A}_q(V)$ and the state $|\Omega_q\rangle$.
\\
We start by observing that, exploiting (\ref{eq:projectorsonGS}) and (\ref{eq:Adecomposition}), we can write
\begin{equation}
\label{eq:decompositiononthestate}
   A\otimes \boldsymbol{1}_{V'}|\Omega\rangle
   =
   \sum_{q\in\mathbb{Z}} A_q \otimes \Pi_{V'}(\bar{q}-q)|\Omega\rangle\,.
\end{equation}
Plugging (\ref{eq:decompositiononthestate}) into (\ref{eq:Modular relation general}) and employing the idempotence of the projectors (\ref{eq:FourierTransform}) and the fact that the terms in (\ref{eq:decompositiononthestate}) with different $q$ are independent from each other, we prove that (\ref{eq:modularrelationfixedq}) is satisfied if
\begin{equation}
\label{eq:SandOmegaatfixedq_new}
S_{\Omega,q}=S_{\Omega}\left[\Pi_V(q)\otimes \Pi_{V'}(\bar{q}-q)\right],
    \qquad\qquad
|\Omega_q\rangle=\left[\Pi_V(q)\otimes \Pi_{V'}(\bar{q}-q)\right]|\Omega\rangle\,.
    \end{equation}
The first equation projects the modular involution $S_{\Omega}$ in the charge sector labeled by $q$, while the state $|\Omega_q\rangle$ belongs to the desired Hilbert space $\mathcal{H}_{V}^{(q)}\otimes\mathcal{H}_{V'}^{(\bar{q}-q)}$, as discussed in Sec.\,\ref{subsec:twistchargesector}.
A further step consists in performing the polar decomposition of $S_{\Omega,q}$, namely
\begin{equation}
\label{eq:SRinvolution}
S_{\Omega,q}=J_{\Omega,q}\left(\Delta_{\Omega,q}\right)^{1/2}\,,
\end{equation}
which is consistent with (\ref{eq:SandOmegaatfixedq_new}) provided that
\begin{equation}
\label{eq:SRmodoperator}
    J_{\Omega,q}=J_{\Omega}\,,
    \qquad\qquad
\Delta_{\Omega,q}=\Delta_{\Omega}\left(\Pi_V(q)\otimes \Pi_{V'}(\bar{q}-q)\right)\,,
\end{equation}
where $\Delta_{\Omega}$ and $J_{\Omega}$ are defined in (\ref{eq:polardecompositiongeneral}).
The last property we need to verify is that $|\Omega_q\rangle\in\mathcal{H}_{V}^{(q)}\otimes\mathcal{H}_{V'}^{(\bar{q}-q)}$ in (\ref{eq:SandOmegaatfixedq_new}) is cyclic and separating for $\mathcal{A}_q(V)$, provided that $|\Omega\rangle\in\mathcal{H}_{\bar{q}}$ is cyclic and separating for $\mathcal{A}(V)$. A sketch of the proof of this fact is reported in Appendix \ref{apx:omegaq_cyclic}.
This feature of $|\Omega_q\rangle$ and the modular relation (\ref{eq:modularrelationfixedq}) allows to regard $\Delta_{\Omega,q}$ defined in (\ref{eq:SRmodoperator}) as a well-defined modular operator. Since it is associated to the algebra $\mathcal{A}_q(V)$ and the state $|\Omega_q\rangle$, this operator encodes the properties of the bipartition $V\cup V'$ in a given charge sector.
A projection onto charge sectors similar to the one in (\ref{eq:SRmodoperator}) can also be achieved for the relative modular operator, which is an important tool given its relation with the relative entropy \cite{Araki:1976zv}. We report and discuss this result in Appendix \ref{apx:relmodflow}.

Since we are working with finite-dimensional algebras, we can employ the expression (\ref{eq:modopRDM0}) for the modular operator in terms of reduced density matrices. Defining
\begin{equation}
\label{eq:SRRDM}
\Pi_V(q) \rho_V\equiv p_V(q)\rho_V(q)\,,\qquad\quad
p_V(q)\equiv \textrm{Tr}\left[\Pi_V(q) \rho_V\right]\,,
\end{equation}
\begin{equation}
\label{eq:SRRDM_complement}
\Pi_{V'}(\bar{q}-q) \rho_{V'}\equiv p_{V'}(\bar{q}-q)\rho_{V'}(\bar{q}-q)\,,
\qquad\qquad
   p_{V'}(\bar{q}-q) \equiv \textrm{Tr}\left[\Pi_{V'}(\bar{q}-q) \rho_{V'}\right]
    \,,
\end{equation}
and using (\ref{eq:SRmodoperator}) and the fact that $\rho_V$ and $\rho_{V'}$ commute with the projectors, we obtain
\begin{equation}
\label{eq:SRmodop_new}
\Delta_{\Omega,q}=\frac{p_V(q)}{p_{V'}(\bar{q}-q)}\rho_V(q)\otimes\left[\rho_{V'}(\bar{q}-q)\right]^{-1}
=\rho_V(q)\otimes\left[\rho_{V'}(\bar{q}-q)\right]^{-1}\,,
\end{equation}
where in the last step we have used that $p_V(q)=p_{V'}(\bar{q}-q)$. This fact can be straightforwardly checked from the definitions (\ref{eq:FourierTransform}), (\ref{eq:SRRDM}) and (\ref{eq:SRRDM_complement}) and can be understood from the interpretation of $p_V(q)$ and $p_{V'}(\bar{q}-q)$ as the probabilities of measuring a charge equal to $q$ and $\bar{q}-q$ in the subsystems $V$ and $V'$ respectively \cite{Xavier:2018kqb}.
Notice that, since we have factored out the trace of $\Pi_{V}(q)\rho_{V}$ and $\Pi_{V'}(\bar{q}-q)\rho_{V'}$, $\rho_V(q)$ and $\rho_{V'}(\bar{q}-q)$ have trace equal to one and can be regarded as proper reduced density matrices restricted to a given charge sector.
The entanglement entropy of these reduced density matrices in different charge sectors is what is known as {\it symmetry-resolved entanglement entropy}, which has recently given rise to an active line of research.

The expressions above allow for the analysis of the modular flow (\ref{eq:modflowRDM}) in the various charge sectors.
Adapting the definition below (\ref{eq:TTtheorem}) to the algebra $\mathcal{A}_q(V)$
and exploiting (\ref{eq:SRmodoperator}) and (\ref{eq:SRmodop_new}), we obtain
\begin{equation}
\label{eq:SRmodflow_new}
\Delta_{\Omega,q}^{\mathrm{i}t}\left(A_q\otimes\boldsymbol{1}_{V'}\right)\Delta_{\Omega,q}^{-\mathrm{i}t}=
 \left[\rho_V(q)\right]^{\mathrm{i}t}
    A_q\left[\rho_V(q)\right]^{-\mathrm{i}t}\otimes\boldsymbol{1}_{V'}
     \equiv
     \sigma_{t,q}(A_q)\otimes\boldsymbol{1}_{V'}
    \,.
\end{equation}
This means that the modular flow (\ref{eq:SRmodflow_new}) in the sector with charge $q$ preserves the corresponding local algebra $\mathcal{A}_q(V)$, as expected from a well-defined modular flow (see \eqref{eq:TTtheorem}).
Using (\ref{eq:modflowRDM}), (\ref{eq:Adecomposition}), (\ref{eq:SRRDM}), (\ref{eq:SRmodflow_new}), the idempotence of the projectors and the fact that $\rho_V(q)$ commutes with $\Pi_V(q)$, we find
\begin{equation}
\label{eq:modflow_decomposition}
\sum_{q\in\mathbb{Z}}  \sigma_{t,q}(A_q)\otimes\boldsymbol{1}_{V'}
=
\sigma_{t}(A)\otimes\boldsymbol{1}_{V'}
\,.
\end{equation}
The decomposition (\ref{eq:modflow_decomposition}) and the analysis above in Sec.\,\ref{subsec:AQFT_srmodflow} are fundamental results of this work, since allow for the interpretation of $\sigma_{t,q}(A_q)\otimes\boldsymbol{1}_{V'}$ as {\it symmetry-resolved modular flow}. Along the same line, a symmetry resolution of the Connes-Radon-Nikodyn flow obtained from the relative modular operator is reported in Appendix \ref{apx:relmodflow}.
%

We can apply the symmetry resolution discussed above also to the modular correlation functions defined in (\ref{eq:modcorrfunc_AQFT}). Exploiting (\ref{eq:modflow_decomposition}), the decomposition (\ref{eq:Adecomposition}) for the operator $B$ and the orthogonality between the projectors onto different charge sectors, we obtain 
\begin{equation}
G_{\textrm{\tiny mod}}(t)
    =
    \sum_{q,q'\in\mathbb{Z}}
    \langle\Omega|B_{q'}\,\sigma_{t,q}(A_q)\otimes\boldsymbol{1}_{V'}|\Omega\rangle
    =\sum_{q\in\mathbb{Z}}
    \langle\Omega|B_{q}\,\sigma_{t,q}(A_q)\otimes\boldsymbol{1}_{V'}|\Omega\rangle
    \,,
\label{eq:SRmodularcorrelations_steps}
\end{equation}
where $A_q\otimes\boldsymbol{1}_{V'}\,,B_q\otimes\boldsymbol{1}_{V'}\in \mathcal{A}_q(V)$ given that $A\otimes\boldsymbol{1}_{V'}\,,B\otimes\boldsymbol{1}_{V'}\in \mathcal{A}(V)$.
Using the properties of the projectors $\Pi_V(q)$, (\ref{eq:projectorsonGS}) and the definition in (\ref{eq:SandOmegaatfixedq_new}),
we observe that
\begin{equation}
\label{eq:SRmodularcorrelations_steps2}
\langle\Omega|B_{q}\,\sigma_{t,q}(A_q)\otimes\boldsymbol{1}_{V'}|\Omega\rangle=\langle\Omega_q|B_{q}\,\sigma_{t,q}(A_q)\otimes\boldsymbol{1}_{V'}|\Omega_q\rangle\,.
\end{equation}
Notice that, recalling the definition (\ref{eq:SRRDM}) of the probability $p_V(q)$,
\be
\label{eq:normOmegaq}
\langle\Omega_q|\Omega_q\rangle=p_V(q)\,,
\ee
namely the state $|\Omega_q\rangle$ is not normalized. This fact must be taken into account in defining a proper symmetry-resolved modular correlation function. Indeed, the decomposition (\ref{eq:SRmodularcorrelations_steps}) can be rewritten as
\begin{equation}
   G_{\textrm{\tiny mod}}(t)=
   \sum_{q\in\mathbb{Z}}
 p_V(q) \frac{\langle\Omega_q|B_{q}\,\sigma_{t,q}(A_q)\otimes\boldsymbol{1}_{V'}|\Omega_q\rangle}{\langle\Omega_q|\Omega_q\rangle} 
  \equiv
    \sum_{q\in\mathbb{Z}}
    p_V(q)G_{\textrm{\tiny mod}}(t,q) \,,
    \label{eq:SRmodularcorrelations}
\end{equation}
where the {\it symmetry-resolved modular correlation function} $G_{\textrm{\tiny mod}}(t,q)$ has been introduced. The decomposition (\ref{eq:SRmodularcorrelations}), which is one of the main results of this manuscript, implies that the knowledge of all the symmetry-resolved modular correlation functions allows to determine the total modular correlation function $G_{\textrm{\tiny mod}}(t)$. 

Exploting (\ref{eq:Adecomposition}), (\ref{eq:SRRDM}), (\ref{eq:SRmodflow_new}) and the fact that $\Pi_V(q)$ commutes with all the operators in $\mathcal{A}(V)$, it is useful to rewrite the left-hand side of (\ref{eq:SRmodularcorrelations_steps2}) as 
\be
\label{eq:modular corr_step1}
\left\langle\Omega|B_q\,\sigma_{t,q}\left(A_q\right)|\Omega\right\rangle
=
\left\langle\Omega|\Pi_V(q)B\,\sigma_{t}\left(A\right)|\Omega\right\rangle
=
\int_{-\pi}^{\pi}\frac{d\alpha}{2\pi}
e^{-\mathrm{i}\alpha q}  \left\langle\Omega|e^{\mathrm{i}\alpha Q_V}B\,\sigma_t\left(A\right)|\Omega\right\rangle
\,,
\ee
where in the last step we have exploited the Fourier representation (\ref{eq:FourierTransform}) of the projector and in all the correlators we have omitted (and we will do it from now on) the tensor product with $\boldsymbol{1}_{V'}$ since it acts trivially on $|\Omega\rangle$. As we will discuss in the forthcoming sections, the expressions in (\ref{eq:modular corr_step1}) are useful for explicit computations of the symmetry-resolved modular correlation function and also allow to determine some features of $G_{\textrm{\tiny mod}}(t,q)$. 
Indeed, by looking at the second expression of (\ref{eq:modular corr_step1}), we can conclude that $G_{\textrm{\tiny mod}}(t,q)$ as function of the modular parameter $t$ is analytic in the strip $-1<\textrm{Im}(t)<0$, according to the discussion in Sec.\,\ref{subsec:TTtheory}. 
Moreover, since $\Pi_V(q)$ is supported on the region $V$, the second expression in (\ref{eq:modular corr_step1}), and therefore $G_{\textrm{\tiny mod}}(t,q)$, satisfies the KMS condition for any value of $q$, namely 
\be
\label{eq:SR_KMS condition}
\left\langle\Omega_q|B_q\,\sigma_{t,q}\left(A_q\right)|\Omega_q\right\rangle
=
\left\langle\Omega_q|\sigma_{t+\mathrm{i},q}\left(A_q\right)\,B_q|\Omega_q\right\rangle
\,,
\ee
which is expected to hold for $\sigma_{t,q}(\cdot)$ to be regarded as a full-fledged modular flow.
\\
Finally, an expression similar to (\ref{eq:modular corr_step1}) can be written down for the probability $p_V(q)$ in (\ref{eq:SRRDM}). Using (\ref{eq:normOmegaq}), (\ref{eq:SandOmegaatfixedq_new}) and (\ref{eq:FourierTransform}), we obtain,
\be
\label{eq:pq_FT}
p_V(q)=
\int_{-\pi}^{\pi}\frac{d\alpha}{2\pi}
e^{-\mathrm{i}\alpha q}  \left\langle\Omega|e^{\mathrm{i}\alpha Q_V}|\Omega\right\rangle\,,
\ee
as first provided in \cite{Xavier:2018kqb}.

\section{Free fermionic theories}
\label{sec:FreeFermions}

As a simple and tractable playground, in this section we consider fermionic Gaussian states, a class which includes also the ground states of free fermionic theories. Considering a  UV-regularized continuum theory for which we can formally consider reduced density matrices, we extend the approach developed in \cite{Erdmenger:2020nop} to the computation of symmetry-resolved modular correlation functions. As an explicit example, we focus on the resolution of the modular correlation function of the charge density in free fermionic theories. 




\subsection{Fermionic Gaussian states}
\label{subsec:FF-ddim}

For defining fermionic field theories, we have to consider $\mathbb{Z}_2$-graded nets of local algebras in order to involve operators which anti-commute among each other when spacelike separated. 
 We implicitly assume this algebraic structure and we refer the interested reader to \cite{Arakibook,Carpi:2007fj} for details.
 In AQFT the field operators in the algebras defining the theory are operator-valued distributions acting on a Hilbert space. This means that the field defined at a given point of the spacetime has to be smeared by a test function, namely a function belonging to the set of Schwartz functions. 
 We can imagine the field algebra introduced at the beginning of Sec.\,\ref{subsec:twistchargesector} to be determined by $\mathbb{Z}_2$-graded local algebras generated by the smeared fermionic fields $\psi$ and $\psi^\dagger$.
In the following, with the aim of performing explicit computations, we exploit the formal analogy with finite-dimensional quantum systems and therefore the smeared fields are replaced by the fields $\psi(x)$ and $\psi^\dagger(x)$ evaluated at points of the spacetime. This analogy can be made precise by regularizing the theory with a UV cutoff $\epsilon$ (as we do, for instance, in Sec.\,\ref{sec:ChiralFF2d}). The quantities computed in the presence of this cutoff are meaningful in QFT if they are well-defined in the limit $\epsilon\to 0$.  This approach has been successfully applied and checked in various cases \cite{Araki:1971id,Casini:2009vk,Hollands:2019hje}.
\\
 The $U(1)$ symmetry group we want to investigate acts on the fields as 
\be
\label{eq:fieldsandsymmetry}
\psi\to e^{\mathrm{i}\alpha}\psi\,,
\qquad\qquad\
\psi^{\dagger}\to e^{-\mathrm{i}\alpha}\psi^{\dagger}
\,.
\ee
Thus, to construct the net of local observable algebras $\mathcal{A}$ as the $U(1)$-invariant part of the field algebra, we consider the union of $\mathcal{A}(V)$ defined as the local algebras generated by $\psi(x)\psi^\dagger(y)$ and $\psi^\dagger(x)\psi(y)$, with $x,y\in V$.

In the following we consider fermionic Gaussian states as an useful playground for determining explicitly modular flow and modular correlation functions.
A Gaussian state of a fermionic field theory in $d$ spatial dimensions is characterized by a reduced density matrix of a generic spatial subsystem $V$ of the form
\be
\label{eq:RDM fermion}
\rho_V=\frac{1}{\mathcal{Z}}\exp
\left[
-
\int_V \mathrm{d}^d x\int_V \mathrm{d}^d y\,
 \colon\psi^\dagger(x)\,k(x,y)\,\psi(y)\colon
\right]\,,
\ee
where the fermionic fields satisfy the canonical anti-commutation relation
\be
\label{eq:CAR}
\{\psi^\dagger(x),\psi(y)\}=\delta(x-y),
\qquad\quad
\{\psi^\dagger(x),\psi^\dagger(y)\}=\{\psi(x),\psi(y)\}=0.
\ee
The constant
$\mathcal{Z}$ is the normalisation that fixes the trace of $\rho_V$ equal to one and the symbol $\colon\;\colon$ denotes the normal ordering.
In this section we consider the theory at a constant time slice and we do not report the dependence of the fermionic fields on the time co-ordinate. Thus, the anti-commutation relations (\ref{eq:CAR}) are at equal times.
The kernel $k(x,y)$, which has to be understood in a distributional sense, is related to the propagator of the theory in the subregion $V$ as follows \cite{Peschel03,Araki:1971id}
\be
\label{eq:kernel_correlator}
e^{-k}=\frac{1-G_V}{G_V},
\ee
where
\be
\label{eq:2pt func def}
G_V(x,y)\equiv\langle\Omega|\psi(x)\psi^\dagger(y)|\Omega\rangle
=\langle\Omega|\psi^\dagger(x)\psi(y)|\Omega\rangle\,,
\qquad
x,y\in V
\,.
\ee
From (\ref{eq:CAR}), the correlator in (\ref{eq:2pt func def}) has the property (in the distributional sense)
\be
\label{eq:CAR_correlator}
G_V(x,y)+G_V(y,x)=\delta(x-y).
\ee
Consider now the {\it charge operator}
\be
\label{eq:chargedef}
Q=\int \colon\psi^{\dagger}(x)\psi(x)\colon \mathrm{d}^d x\,,
\ee
which generates the $U(1)$ symmetry of the fermionic theory implemented on the fields as shown in (\ref{eq:fieldsandsymmetry}). 
This operator can be restricted to the subsystem $V$ as
\be
\label{eq:reducedchargedef}
Q_V=\int_V \colon\psi^{\dagger}(x)\psi(x)\colon \mathrm{d}^d x\,.
\ee
The restriction of the charge density (\ref{eq:reducedchargedef}) to the subsystem $V$ is not well-defined in algebraic QFT, but we can make sense of it in our setup since we are exploiting the analogy with the finite-dimensional case, where the factorization of the Hilbert space and the reduced density matrices are meaningful.
We observe that $Q_V$ commute with $\rho_V$, as can be verified using (\ref{eq:RDM fermion}) and (\ref{eq:CAR}).
It is also straightforward to check that $[Q_V,\psi(x)]\neq 0$ and $[Q_V,\psi^\dagger(x)]\neq 0$ for any $x\in V$, but $[Q_V,\psi^\dagger(x)\psi(x)]= 0$. 
Thus, in order to discuss the symmetry resolution of the modular flow and of the modular correlation function a suitable operator to consider is the charge density $\colon\psi^{\dagger}(x)\psi(x)\colon$, with $x\in V$. 
In particular, we are interested in
the following modular correlation function
\be
\label{eq:modcorrfunc_FF}
G^{({\rm c})}_{\textrm{\tiny mod}}(x,y;t)
\equiv
\langle\Omega|\colon \psi^\dagger(x) \psi(x) \colon \sigma_t\left(\!\colon \psi^\dagger(y) \psi(y) \colon\!\right)|\Omega\rangle,
\quad\quad
-1< {\rm Im}(t)< 0\,,
\ee
where the modular flow $\sigma_t(\cdot)$ is defined in (\ref{eq:modflowRDM}) and the superscript ${\rm c}$ refers to the fact that we are evolving the charge density operator through the modular flow. This modular correlation function, which has been studied in various cases of interest in the literature \cite{Hollands:2019hje,Mintchev:2020jhc,Mintchev:2020uom}, is analyitic when $-1< {\rm Im}(t)< 0 $ and can be continued to the strip $0< {\rm Im}(t)< 1 $ through the KMS condition
\be
\label{eq:KMS FF}
\langle\Omega|\colon \psi^\dagger(x) \psi(x) \colon \sigma_t\left(\!\colon \psi^\dagger(y) \psi(y) \colon\!\right)|\Omega\rangle=
\langle\Omega|\sigma_{t+\mathrm{i}}\left(\!\colon \psi^\dagger(y) \psi(y) \colon\!\right)\colon \psi^\dagger(x) \psi(x) \colon |\Omega\rangle\,.
\ee
 The analytic continuation of $G^{({\rm c})}_{\textrm{\tiny mod}}$ coming from (\ref{eq:KMS FF}) implies
\be
G^{({\rm c})}_{\textrm{\tiny mod}}(x,y;t)=G^{({\rm c})}_{\textrm{\tiny mod}}(x,y;t+\mathrm{i})\,,
\qquad\qquad
-1< {\rm Im}(t)< 0
\,.
\ee
In other words, the modular correlation function of the charge operator satisfies periodic boundary conditions along imaginary values of the modular parameter $t$, as opposed to the antiperiodic boundary conditions occurring for a single fermionic field \cite{Erdmenger:2020nop}. This is due to the bosonic nature of the charge density we are considering.

\subsection{Modular correlation function of fermionic Gaussian states}
\label{subsec:modflow_MCF_generaddim}

In \cite{Erdmenger:2020nop} it has been found that the modular flow of the field $\psi^\dagger(x)$, with $x\in V$, can be written as
\be
\label{eq:mod flow FF_kernel}
\sigma_t\left(\psi^\dagger(x)\right)=
\int_V \psi^{\dagger}(y)\Sigma_t(y,x) \mathrm{d}^d y\,,
\qquad\qquad
\Sigma_t=\left[\frac{1-G_V}{G_V}\right]^{{\rm i}t}\,.
\ee
The explicit expression of $\Sigma_t$ as function of $x,y$ and $t$ can be obtained in some cases by using the resolvent method (for details on this method see \cite{Casini:2009vk,Klich:2015ina,Fries:2019ozf,Blanco:2019xwi}).
The correlation function along the modular flow (\ref{eq:mod flow FF_kernel}) can be written, for $-1< {\rm Im}(t)< 0 $ and $x,y\in V$, as
\be
\label{eq:mod corrfunc FF_kernel}
G^{\textrm{\tiny (f)}}_{\textrm{\tiny mod}}(x,y;t)\equiv\langle\Omega| \psi(x)  \sigma_t\left( \psi^\dagger(y) \right)|\Omega\rangle=
\left(
G_V\left[\frac{1-G_V}{G_V}\right]^{{\rm i} t}
\right)(x,y)\,.
\ee
The superscript $\textrm{f}$ in (\ref{eq:mod corrfunc FF_kernel}) refers to the fact that we are computing the modular correlation function of a single fermionic field, differently from (\ref{eq:modcorrfunc_FF}) where the charge density is considered.
Given that (\ref{eq:mod corrfunc FF_kernel}) is written as a functional of the restricted correlator $G_V$, it can also be computed using the resolvent, whenever it is known. The modular correlator $G^{\textrm{\tiny (f)}}_{\textrm{\tiny mod}}(x,y;t)$ satisfies the KMS condition \cite{Haagbook,Haag:1967sg,Hollands:2019hje}
\be
\label{eq:KMS single FF}
\langle\Omega| \psi(x)  \sigma_t\left(\psi^\dagger(y) \right)|\Omega\rangle=-
\langle\Omega|\sigma_{t+\mathrm{i}}\left( \psi^\dagger(y) \right)\psi(x) |\Omega\rangle,
\ee
or, equivalently,
\be
G^{({\rm f})}_{\textrm{\tiny mod}}(x,y;t)=-G^{({\rm f})}_{\textrm{\tiny mod}}(x,y;t+\mathrm{i}),
\ee
where the minus sign comes from the fermionic statistics of the fields evolved through the modular flow.

We now turn to the modular correlation function (\ref{eq:modcorrfunc_FF}) of the charge density and we adapt the formalism developed in \cite{Erdmenger:2020nop} to its computation. For this purpose, we preliminarily observe that 
\be
\label{eq:normal ordering}
\sigma_t\left(\!
\colon\psi^{\dagger}(y)\psi(y)\colon\!
\right)
=\,
\colon\sigma_t\left(
\psi^{\dagger}(y)\psi(y)
\right)
\colon\,.
\ee
This means we can consider only $\sigma_t\left(\psi^{\dagger}(y)\psi(y)\right)$, observing that
\be
\sigma_t\left( \psi^\dagger(y) \psi(y) \right)
=
\rho_V^{\mathrm{i}t} \psi^\dagger(y) \psi(y) \rho_V^{-\mathrm{i}t}
=
\sigma_t\big(\psi^\dagger(y)\big)\sigma_t\big( \psi^\dagger(y) \big)^\dagger\,,
\label{ModFlowSimple_1}
\ee
where in the first step we have exploited (\ref{eq:modflowRDM}). Now we can use the results for the modular flow of a single field and, plugging (\ref{eq:mod flow FF_kernel}) into (\ref{ModFlowSimple_1}), we obtain
\be
\label{ModFlowSimple_2}
\sigma_t\left(\!\colon \psi^\dagger(y) \psi(y) \colon\!\right)=
\colon\int_V\mathrm{d}^d x \psi^\dagger(x)\Sigma_t(x,y)\int_V\mathrm{d}^d z\psi(z)\Sigma^*_t(z,y)\colon.
\ee
We further manipulate this expression to get rid of the complex conjugation of $\Sigma_t$ in the second integral of the right-hand side.
Employing the definition (\ref{eq:mod flow FF_kernel}), the complex conjugation of $\Sigma_t$ reads
\be
\label{eq:SigmaStar}
\Sigma^*_t=\left[\frac{1-G_V^*}{G_V^*}\right]^{-\mathrm{i}t}.
\ee
Moreover, given the general definition of the propagator (\ref{eq:2pt func def}), we have
\bea
\label{eq:correlator conjugate 2}
\label{correlator conjugate}
G_V^*(x,y)&=&\langle\Omega|\psi(x)\psi^\dagger(y)|\Omega\rangle^*=\langle\Omega|\psi(y)\psi^\dagger(x)|\Omega\rangle=G_V(y,x)
\\
&=&
\nonumber
\delta(x-y)-G_V(x,y)=[1-G_V](x,y)\,,
\eea
where in the last step we have exploited (\ref{eq:CAR_correlator}).
Plugging (\ref{eq:correlator conjugate 2}) into (\ref{eq:SigmaStar}), we get
\be
\label{eq:SigmaConjugate2}
\Sigma^*_t(x,y)=\Sigma_t(x,y),
\ee
and therefore 
\be
\label{eq:ModFlowSimple_4}
\sigma_t\left(\!\colon \psi^\dagger(y) \psi(y) \colon\!\right)=
\colon\int_V\mathrm{d}^d x \psi^\dagger(x)\Sigma_t(x,y)\int_V\mathrm{d}^d z\psi(z)\Sigma_{t}(z,y)\colon,
\ee
which shows that the modular flow of the charge density is given by the product of the modular flows of the single fields.
As for the modular correlation function,
using (\ref{eq:ModFlowSimple_4}) and (\ref{eq:normal ordering}) in (\ref{eq:modcorrfunc_FF}), we obtain
\be
\label{FullmodularcorrelationFF}
G^{({\rm c})}_{\textrm{\tiny mod}}(x,y;t)
=
  \int_V
\mathrm{d}^d x_1\mathrm{d}^d x_2
\Sigma_t (x_1,y)\Sigma_t(x_2,y)F(x,x_1,x_2)\,,
\ee
where we have defined
\be
\label{FuncF_def}
F(x,x_1,x_2)\equiv\left\langle\Omega\left|\colon\psi^{\dagger}(x)\psi(x)\colon
  \colon\psi^{\dagger}(x_1)\psi(x_2)\colon
  \right|\Omega\right\rangle\,.
\ee
The last obstacle to obtain $G^{({\rm c})}_{\textrm{\tiny mod}}(x,y;t)$ is therefore computing the correlation function in (\ref{FuncF_def}).
This can be evaluated using Wick's theorem and the result is simply
\be
\label{Wickusing_v0}
F(x,x_1,x_2)=G_V(x,x_1)G_V(x,x_2)\,.
\ee
Notice that we can exploit the Wick's theorem given that we are dealing with fermionic Gaussian states.
Plugging (\ref{Wickusing_v0}) back into (\ref{FullmodularcorrelationFF}), we obtain
\be
\label{FullmodularcorrelationFF_v2}
G^{\textrm{\tiny (c)}}_{\textrm{\tiny mod}}(x,y;t)
=
 \int_V
\mathrm{d}^d x_1\mathrm{d}^d x_2
G_V(x,x_1)\Sigma_t (x_1,y)G_V(x,x_2)\Sigma_t (x_2,y)
=
\left[G^{\textrm{\tiny (f)}}_{\textrm{\tiny mod}}(x,y;t)\right]^2\,,
\ee
where the factorization of the two integrals in the second term of (\ref{FullmodularcorrelationFF_v2}) allows to identify on the right-hand side the square of the modular correlation function (\ref{eq:mod corrfunc FF_kernel}) of a single fermionic field. 
Thus, the knowledge of $G^{\textrm{\tiny (f)}}_{\textrm{\tiny mod}}(x,y;t)$, immediately allows  to determine $G^{\textrm{\tiny (c)}}_{\textrm{\tiny mod}}(x,y;t)$.
Moreover, (\ref{FullmodularcorrelationFF_v2}) automatically implies that (\ref{eq:KMS FF}) is satisfied because of the KMS condition (\ref{eq:KMS single FF}) for the modular correlation function of a single field and the cancellation of the two minus signs.
Notice that the relation (\ref{FullmodularcorrelationFF_v2}) was already found in \cite{Hollands:2019hje, Mintchev:2020jhc}. The computation reported above allows to retrieve this result by adapting the approach developed in \cite{Erdmenger:2020nop} to the modular correlation function of the charge density. This is one of the main achievements in this work, namely a first application of the aforementioned formalism to obtain the modular flow of composite operators and their flowed correlation functions.

The expression (\ref{ModFlowSimple_2}) for the modular flow can be also exploited as starting point for determining the symmetry resolution of modular correlation function (\ref{FullmodularcorrelationFF}). Adapting the definition (\ref{eq:SRmodularcorrelations}) to the charge denstity operator $\colon\psi^\dagger\psi\colon$ and taking (\ref{eq:normOmegaq}), (\ref{eq:modular corr_step1}) and (\ref{eq:reducedchargedef}) into account, we want to compute
\be
\label{eq:SR_modcorr FF}
G^{\textrm{\tiny (c)}}_{\textrm{\tiny mod}}(x,y;t,q)
\equiv
\int_{-\pi}^{\pi}\frac{d\alpha}{2\pi} \frac{e^{-\mathrm{i}\alpha q}}{p_V(q)}
\left\langle\Omega\left|e^{\mathrm{i}\alpha \int_V :\psi^{\dagger}(x')\psi(x')\colon \mathrm{d}
^d x'}\colon\psi^{\dagger}(x)\psi(x)::\sigma_t\left(\psi^{\dagger}(y)\psi(y)\right):
  \right|\Omega\right\rangle
  \,,
\ee
where, for fermionic Gaussian states, the probability $p_V(q)$ introduced in (\ref{eq:SRRDM})
reads
\be
\label{eq:prob_FT}
p_V(q)=
\int_{-\pi}^{\pi}\frac{d\alpha}{2\pi}e^{-\mathrm{i}\alpha q}
\langle\Omega|e^{\mathrm{i}\alpha \int_V \colon\psi^{\dagger}(x')\psi(x')\colon \mathrm{d}
^d x'}|\Omega \rangle\,.
\ee
The result (\ref{eq:SR_modcorr FF}) moves the difficulty in obtaining the symmetry-resolved correlation function to the computation of the correlation function on the right-hand side. 
We can rewrite it in light of the method discussed above in this section and, employing (\ref{eq:ModFlowSimple_4}) in (\ref{eq:SR_modcorr FF}), we find
\be
\label{eq:SRmodcorrfunc_FT}
G^{\textrm{\tiny (c)}}_{\textrm{\tiny mod}}(x,y;t,q)=
\int_V
\mathrm{d}^dx_1\mathrm{d}^dx_2
\Sigma_t (x_1,y)\Sigma_t (x_2,y)
\int_{-\pi}^{\pi}\frac{d\alpha}{2\pi} 
\frac{e^{-\mathrm{i}\alpha q}}{p_V(q)}F(x,x_1,x_2;\alpha),
\ee
where
\be
\label{eq:functionFalpha_def}
F(x,x_1,x_2;\alpha)\equiv \left\langle\Omega\left|e^{\mathrm{i}\alpha \int_V \colon\psi^{\dagger}(x')\psi(x')\colon \mathrm{d}^dx'}\colon\psi^{\dagger}(x)\psi(x)\colon
  \colon\psi^{\dagger}(x_1)\psi(x_2)\colon
  \right|\Omega\right\rangle.
\ee
This result is an important finding of this manuscript, since it provides a formula for computing the symmetry-resolved modular correlation function of the charge density in fermionic Gaussian states.
In Sec.\,\ref{sec:ChiralFF2d}\,, the function $F(x,x_1,x_2;\alpha)$ and the resolved modular correlation function (\ref{eq:SRmodcorrfunc_FT}) are explicitly evaluated for the example of a $1+1$-dimensional free massless Dirac field theory.
Notice that $F(x,x_1,x_2;\alpha=0)=F(x,x_1,x_2)$, with $F(x,x_1,x_2)$ defined in (\ref{FuncF_def}). Thus, since in the distributional sense 
\be
\label{eq:delta alpha rep}
\sum_{q\in \mathbb{Z}}\frac{e^{-\mathrm{i}\alpha q}}{2\pi}=\delta(\alpha),
\ee
we have
\be
\label{eq:resum_SRMCF_FF}
\sum_{q\in \mathbb{Z}}p_V(q) G^{\textrm{\tiny (c)}}_{\textrm{\tiny mod}}(x,y;t,q)= G^{\textrm{\tiny (c)}}_{\textrm{\tiny mod}}(x,y;t),
\ee
consistently with the decomposition in (\ref{eq:SRmodularcorrelations}).
The discussion and the results of this section can be straightforwardly generalized to the case of a Gaussian states of $N$ non-interacting fermions, as done in Appendix \ref{apx:NFreeFermions}.

\section{\texorpdfstring{$1+1$}{1+1}-dimensional free massless Dirac field theory}
\label{sec:ChiralFF2d}

In this section, we apply the discussion from the previous pages to the ground state of a $1+1$-dimensional massless Dirac field theory bipartite into a region made by $p$ disjoint intervals and its complement. Through the computation outlined in Sec.\,\ref{sec:FreeFermions}, we recover the known result about the modular correlation function of the charge density and we compute for the first time the corresponding symmetry-resolved modular correlation function. Remarkably, the latter quantity exhibits independence of the $U(1)$ charge sector at leading order in the UV cutoff.

\subsection{General properties of the theory}
\label{subsec:ChiralFF2d_def}

Consider a $1+1$-dimensional free massless Dirac theory on an infinite line, described by the action
\be
\mathcal{S}_\textrm{\tiny D}=\int\textrm{d}\tau\textrm{d}x\, \textrm{i}\bar{\boldsymbol{\psi}}\left(\gamma^0\partial_0+\gamma^1\partial_1\right)\boldsymbol{\psi},
\ee
where $\boldsymbol{\psi}=(\psi_+,\psi_-)^\textrm{t}$ is the two-component Dirac field, the matrices $\gamma^\mu$ are given by the Pauli matrices as $\gamma^0=\sigma_1$ and $\gamma^1=\textrm{i}\sigma_2$ and $\bar{\boldsymbol{\psi}}=\boldsymbol{\psi}^\dagger\gamma^0$. 
By introducing the co-ordinates $x^{\pm}=\tau\pm x$, we can rewrite the action as
\be
\label{eq:actionDirac}
\mathcal{S}_\textrm{\tiny D}=\mathcal{S}_\textrm{\tiny D}^{(+)}+\mathcal{S}_\textrm{\tiny D}^{(-)},
\qquad\qquad
\mathcal{S}_\textrm{\tiny D}^{(\pm)}=\int\textrm{d}x^{+}\textrm{d}x^{-}\;\textrm{i}\psi_\pm^\dagger\partial_\mp\psi_\pm.
\ee
The equations of motion of the two fields read
\be
\partial_+ \psi_-=0\,,
\qquad
\partial_- \psi_+=0\,.
\ee
This means that $\psi_+$ is a function of the co-ordinate $x^+$ only and $\psi_-$ of the co-ordinate $x^-$, namely $\psi_\pm=\psi_\pm(\tau\pm x)$.
The two actions have independent $U(1)$ symmetries, namely are invariant under the phase transformations
\be
\label{eq:chiralsymmetry}
\psi_\pm\to e^{\mathrm{i}\alpha_\pm}\psi_\pm,
\ee
where the $U(1)$ parameters $\alpha_\pm$ are in general different.
By Noether's theorem, this leads to the conservation of two charges $Q_\pm$ given by
\be
\label{eq:chiral charge densities}
Q_+=\int \textrm{d}x \colon\psi^\dagger_+ \psi_+\colon
\equiv
\int \textrm{d}x j_+(\tau+x)\,,
\qquad
Q_-=\int \textrm{d}x
\colon\psi^\dagger_- \psi_-\colon
\equiv
\int \textrm{d}x j_-(\tau-x)\,,
\ee 
where we have introduced the notation $j_\pm$ for the charge density of the two chiral fields $\psi_+$ and $\psi_-$ and the integrals run over slices with fixed co-ordinate $\tau$.
The conserved quantities in (\ref{eq:chiral charge densities}) are interpreted as the number of left-movers ($Q_+$) and right-movers ($Q_-$) in the theory and are separately conserved (this does not happen in presence of a mass term in (\ref{eq:actionDirac})).
The total charge 
\be
\label{eq:def_total charge}
Q\equiv Q_++Q_-
=
\int \textrm{d}x j(\tau, x)\,,
\qquad
j(\tau, x)
\equiv
j_+(\tau+ x)+j_-(\tau- x)\,,
\ee
generates the $U(1)$ symmetry realized on the Dirac field as $\boldsymbol{\psi}\to e^{\mathrm{i}\alpha}\boldsymbol{\psi}$, corresponding to $\alpha_+=\alpha_-$ in (\ref{eq:chiralsymmetry}) for the chiral fields.

From now on, we consider the theory on a constant time slice and therefore we will omit the dependence of all the fields on the co-ordinate $\tau$.
The fields $\psi_+$ and $\psi_-$ satisfy canonical anti-commutation relations at equal times
\be
\label{eq:CAR_2d}
\{\psi_+^\dagger(x),\psi_+(y)\}=\{\psi^\dagger_-(x),\psi_-(y)\}=\delta(x-y)\,,
\ee
\be
\{\psi_+(x),\psi_+(y)\}=
\{\psi_+^\dagger(x),\psi^\dagger_+(y)\}=
\{\psi_-(x),\psi_-(y)\}=
\{\psi^\dagger_-(x),\psi^\dagger_-(y)\}=0\,,
\ee
and are mutually anticommuting.
Since the action $\mathcal{S}_\textrm{\tiny D}$ in (\ref{eq:actionDirac}) is quadratic in the fields, the ground state $|\Omega\rangle$ of this theory is a Gaussian state and all the correlators on $|\Omega\rangle$ can be computed via Wick's theorem in terms of the two-point functions. When the fields are evaluated at the same time co-ordinate, the two-point functions read
\be
\label{eq:2p_corr_chiral}
G^{\textrm{\tiny (+)}}(x,y)\equiv
\langle\Omega|\psi_+(x)\psi_+^\dagger(y)|\Omega\rangle
=\langle\Omega|\psi_+^\dagger(x)\psi_+(y)|\Omega\rangle=
\lim_{\epsilon\to 0^+}
\frac{1}{2\pi\mathrm{i}}\frac{1}{x-y-\mathrm{i}\epsilon}\,,
\ee
\be
\label{eq:2d_corr_antichiral}
G^{\textrm{\tiny $(-)$}}(x,y)\equiv
\langle\Omega|\psi_-(x)\psi_-^\dagger(y)|\Omega\rangle
=\langle\Omega|\psi_-^\dagger(x)\psi_-(y)|\Omega\rangle=
\lim_{\epsilon\to 0^+}
-\frac{1}{2\pi\mathrm{i}}\frac{1}{x-y+\mathrm{i}\epsilon}\,.
\ee
Notice that the parameter $\epsilon$ acts as an ultraviolet (UV) cutoff. From now on, even if not explicitly reported, the limit $\epsilon\to 0^+$ is implied.

In the following, we study the properties of the modular flow and the modular correlation functions of this theory in its ground state and we discuss their symmetry resolution. Notice that the ground state is an eigenstate of the charge (\ref{eq:def_total charge}) with eigenvalue equal to zero and therefore it belongs to the subspace $\mathcal{H}_0$, according to the decomposition (\ref{eq:decHS}).
Given that the two actions $\mathcal{S}^{(\pm)}_\mathrm{\tiny D}$  in (\ref{eq:actionDirac}) are decoupled, the ground state of the theory factorizes into a left and a right part and a similar decomposition occurs for the density matrix of the ground state reduced to any spatial subsystem $V$  on a constant time slice. Explicitly, we have
\be
\label{eq:factorization}
\rho_{V}=\rho_V^{\textrm{\tiny $(+)$}}\otimes \rho_V^{\textrm{\tiny $(-)$}}\,.
\ee
 The factorization in (\ref{eq:factorization}) implies a factorization also for the modular flow. In particular, given $O^{(\pm)}$ written as products of fields $\psi_{\pm}$ localized in $V$, exploiting (\ref{eq:modflowRDM}) we can write the modular flow as
\be
\label{eq:factor_modflow}
\hspace{1cm}
\rho_{V}^{\mathrm{i}t}O^{\textrm{\tiny $(+)$}}O^{\textrm{\tiny $(-)$}}\rho_{V}^{-\mathrm{i}t}=
\sigma_t\big(O^{\textrm{\tiny $(+)$}}\big)\otimes\bar{\sigma}_t\big(O^{\textrm{\tiny $(-)$}}\big)\,,
\ee
\be
\sigma_t\big(O^{\textrm{\tiny $(+)$}}\big)
\equiv
\big[\rho^{\textrm{\tiny $(+)$}}_{V}\big]^{\mathrm{i}t}O^{\textrm{\tiny $(+)$}}\big[\rho^{\textrm{\tiny $(+)$}}_{V}\big]^{-\mathrm{i}t}\,,
\qquad\quad
\bar{\sigma}_t(O^{\textrm{\tiny $(-)$}})
\equiv
\big[\rho^{\textrm{\tiny $(-)$}}_{V}\big]^{\mathrm{i}t}O^{\textrm{\tiny $(-)$}}\big[\rho^{\textrm{\tiny $(-)$}}_{V}\big]^{-\mathrm{i}t}\,.
\ee
Another consequence of the factorization of the ground state is that all the correlation functions, including the modular correlation functions, also factorize into a term involving only the fields $\psi_+$ and a term only involving $\psi_-$.

As an explicit example, we assume $V$ to be made up of $p$ disjoint intervals. More precisely we define, $V\equiv\bigcup_{j=1}^p [a_j,b_j]$.
For this bipartition, in the next subsection we compute the modular flow and the modular correlation function at equal times of the charge density operator $j(x)\equiv j_+(x)+j_-(x)$, with $x\in V$, that read
\be
\rho_V^{\mathrm{i}t}j(x)\rho_V^{-\mathrm{i}t}
=
\sigma_t\left(j_+(x)\right)
+\bar{\sigma}_t\left(j_-(x)\right)
\,,
\label{eq:modflow_totaldensity}
\ee
and 
\be
G^{\textrm{\tiny D}}_{\textrm{\tiny mod}}
(t)
\equiv
\langle\Omega| j(x)  \rho_{V}^{\mathrm{i}t}j(y)\rho_{V}^{-\mathrm{i}t}|\Omega\rangle
=G^{\textrm{\tiny (c,+)}}_{\textrm{\tiny mod}}(x,y;t)
+
G^{\textrm{\tiny (c,$-$)}}_{\textrm{\tiny mod}}(x,y;t)
\,,
\label{eq:modcorr_totaldensity}
\ee
where
\be
\label{eq:modcorr_chiraldensity}
G^{\textrm{\tiny (c,+)}}_{\textrm{\tiny mod}}(x,y;t)\equiv
\langle\Omega| j_+(x)  \sigma_t\left(j_+(y) \right)|\Omega\rangle
\,,\qquad
G^{\textrm{\tiny (c,$-$)}}_{\textrm{\tiny mod}}(x,y;t)
\equiv
\langle\Omega| j_-(x)  \bar{\sigma}_t\left(j_-(y) \right)|\Omega\rangle\,.
\ee
Notice that the modular correlator $G^{\textrm{\tiny D}}_{\textrm{\tiny mod}}
(t)$ in 
(\ref{eq:modcorr_totaldensity}) is also a function of the co-ordinates $x$ and $y$, but we omit to specify this dependence to avoid clutter.   
The expressions (\ref{eq:modflow_totaldensity}) and (\ref{eq:modcorr_totaldensity}) imply that we can compute separately the modular flow $\sigma_t$ and $\bar{\sigma}_t$ and the corresponding modular correlation functions. 
Since both the left and the right part of the ground state of the Dirac field theory are Gaussian states, the modular flows and the modular correlation functions can be computed by applying the machinery reviewed in Sec.\,\ref{sec:FreeFermions} with $d=1$ (unless otherwise specified, for the rest of this section we fix $d=1$). 
For obtaining $\sigma_t$ and $G^{\textrm{\tiny (c,+)}}_{\textrm{\tiny mod}}(x,y;t)$, we have to consider $G^{\textrm{\tiny (+)}}_V(x,y)\equiv G^{\textrm{\tiny (+)}}(x,y)|_{x,y\in V} $ as two-point correlator restricted to the subsystem, while, for $\bar{\sigma}_t$ and $G^{\textrm{\tiny (c,$-$)}}_{\textrm{\tiny mod}}(x,y;t)$, the restricted correlator $G^{\textrm{\tiny ($-$)}}_{V}(x,y)\equiv G^{\textrm{\tiny ($-$)}}(x,y)|_{x,y\in V} $ must be taken into account. 
In what follows we provide some computational details for $\sigma_t$ and $G^{\textrm{\tiny (c,+)}}_{\textrm{\tiny mod}}(x,y;t)
$, while we only report the result, if necessary, for $\bar{\sigma}_t$ and $G^{\textrm{\tiny (c,$-$)}}_{\textrm{\tiny mod}}(x,y;t)$.

\subsection{Modular flow and modular correlation functions}
\label{subsec:modflowandcorrelator-FF_noSR}

Exploiting the approach reviewed in Sec.\,\ref{sec:FreeFermions}, the modular flow for the chiral field $\psi_+$ can be written as
\be
\label{eq:modflow_chiraldagger}
\sigma_t\left(\psi_+^\dagger(x)\right)=
\int_V \psi_+^{\dagger}(y)\Sigma_t(y,x) \mathrm{d} y\,,\qquad
\sigma_t\left(\psi_+(x)\right)=
\int_V \psi_+(y)\Sigma_t(y,x) \mathrm{d} y\,.
\ee
 In \cite{Erdmenger:2020nop} the kernel $\Sigma_t(x,y)$ has been explicitly computed exploiting the resolvent method. It reads
\be
\label{eq:knownSigmat}
\Sigma_t(x,y)=-2\mathrm{i}G^{\textrm{\tiny (+)}}_V(x,y)\sinh(\pi t)\delta\left(t+\tilde{t}(x,y)\right)\,,
\ee
where we have defined
\be
\label{eq:tildet_Def}
\tilde{t}(x,y)=Z(x)-Z(y)\,,
\qquad\qquad
Z(x)=\frac{1}{2\pi}\log
\left[
-\prod_{j=1}^p
\frac{a_j-x}{b_j-x}
\right]\,,
\ee
and $G^{\textrm{\tiny (+)}}_V(x,y)$ is given in (\ref{eq:2p_corr_chiral}).
Plugging (\ref{eq:knownSigmat}) into (\ref{eq:modflow_chiraldagger}), we recover the modular flow for the chiral field first obtained in \cite{Casini:2009vk} 
\be
\label{eq:modflow_sumchiraldagger}
\sigma_t\left(\psi_+^\dagger(x)\right)=-2\mathrm{i}\sinh(\pi t)\sum_{l=1}^p
\frac{G^{\textrm{\tiny (+)}}_V\left( x_\ell,x\right)}{Z'\left(x_\ell\right)}\psi_+^\dagger\left(x_\ell\right)\,,
\quad
Z'(x)=\frac{1}{2\pi}\sum_{j=1}^p\frac{b_j-a_j}{(b_j-x)(x-a_j)}\,,
\ee
where $x_\ell$ are solutions of
\be
\label{eq:solution_xell}
t+\tilde{t}(x_\ell,x)=t+Z(x_\ell)-Z(x)=0\,.
\ee
The same expression in (\ref{eq:modflow_sumchiraldagger}) can be obtained for $\psi_+(x)$.
Notice that $Z$ in (\ref{eq:tildet_Def}) is a monotonic function of its argument in each interval of $V$ (and therefore we do not need to consider the absolute value in the denominator of (\ref{eq:modflow_sumchiraldagger})).
Let us comment on what happens to (\ref{eq:modflow_sumchiraldagger}) in the limit $t\to 0$. The prefactor $\sinh(\pi t)$ vanishes linearly in $t$ and therefore the only non-zero contribution comes from the {\it local} solution $x_\ell$, namely the solution of (\ref{eq:solution_xell}) which belongs to the same interval as $x$ and such that $x_\ell\to x$ as $t\to 0$. Since this happens, the zero at the denominator of $G^{\textrm{\tiny (+)}}_V(x_\ell,x)$ cancels the zero of the prefactor and we obtain $\sigma_t\left(\psi_+^\dagger(x)\right)\to \psi_+^\dagger(x) $ as $t\to 0$, which is consistent with the expectation $\Sigma_{t=0}(x,y)=\delta(x-y)$.
The same discussion holds for modular flow of the field $\psi_+(x)$.

Once the modular flows of $\psi_+^\dagger$ and $\psi_+$ are known, using (\ref{eq:ModFlowSimple_4}), we can obtain the modular flow of the charge density $j_+$ introduced in (\ref{eq:chiral charge densities}). It reads
\be
\label{eq:Mod_flow_chargechiralFF}
\sigma_t
\big(
j_+(x)
\big)
=
-4 \left[
\sinh(\pi t)
\right]^2
\sum_{\ell,\ell'=1}^p
\frac{G^{(2\textrm{d})}_V\left( x_\ell,x\right)}{Z'\left(x_\ell\right)}
\frac{G^{(2\textrm{d})}_V\left( x_{\ell'},x\right)}{Z'\left(x_{\ell'}\right)}
\colon\psi_+^\dagger\left(x_\ell\right) \psi_+\left(x_{\ell'}\right)\colon\,.
\ee
To avoid misunderstanding, we stress that this modular flow is the one generated by the modular operator defined on the algebras of fermions discussed in Sec.\,\ref{subsec:FF-ddim}. This is different from the modular flow defined on the subnet of algebras generated by the operators $j_+$ at different spacetime points. The latter is not considered in this manuscript and we simply refer to (\ref{eq:Mod_flow_chargechiralFF}) as modular flow of the charge density.
Following the same argument reported below (\ref{eq:solution_xell}), we can check that, as expected,
$
\sigma_{t=0}
\big(
j_+(x)
\big)
=
j_+(x)\,.
$
Let us comment on the structure of modular flow (\ref{eq:Mod_flow_chargechiralFF}), which is richer than the one of a single chiral field discussed in \cite{Erdmenger:2020nop}.
When $p=2$, i.e. $V=[a_1,b_1]\cup[a_2,b_2]$,
the equation (\ref{eq:solution_xell}) can be solved yielding two solutions, whose expressions are not illuminating and worthwhile to be reported.
What matters is that one of the two solutions lies in the same interval as $x$, while the second one lies in the other interval. A sketch of the solutions in this case can be found in Fig.\,4 of \cite{Erdmenger:2020nop}.
Differently from the case discussed in \cite{Erdmenger:2020nop}, in the final expression of the modular flow (\ref{eq:Mod_flow_chargechiralFF}) pairs of solutions are involved. This leads to three kinds of terms:
\begin{itemize}
\item one term coupling $x$ with two points in the same interval ({\it local term});
\item one term coupling $x$ with two points in the other interval ({\it doubly bi-local term});
\item two terms, one coupling $x$ with a point in the same interval and the other coupling $x$ with a point in the different interval ({\it bi-local terms}).
\end{itemize}  
The generalization to the case with $p>2$ intervals is straightforward. Regardless their explicit expressions, one can find $p$ solutions of (\ref{eq:solution_xell}), each of them localized in one of the intervals which $V$ is made up of. 
In the modular flow of $p$ disjoint intervals, according to the previous classification, we find a unique local term, $2p-2 $ bi-local terms and $(p-1)^2 $ doubly bi-local terms (for consistency, notice that $1+2p-2+(p-1)^2=p^2$, as the number of terms in (\ref{eq:Mod_flow_chargechiralFF}) when $V$ is made by $p$ intervals).
\\
The modular flow of the charge density $j_-(x)$ is 
achieved by adapting the calculation reported above.
First, the modular flow for the single chiral field $\psi_-(x)$ must be computed. The result is achieved by slightly modifying the procedure
of \cite{Erdmenger:2020nop} and
it is given by
(\ref{eq:Mod_flow_chargechiralFF}) with the change of sign $t\to -t$ of the modular parameter \cite{Casini:2009vk}. This is somehow expected if we think to the case of the Rindler space (retrieved from our computation when $p=1$, $a_1=0$ and $b_1\to\infty$), where the modular flow is generated by the boost generator \cite{Bisognano:1975ih,Bisognano:1976za}. This change of sign must also be applied to obtain $\bar{\sigma}_t(j_-(x))$ from $\sigma_t(j_+(x))$ in (\ref{eq:Mod_flow_chargechiralFF}).
Thus, the modular flow $\bar{\sigma}_t(j_-(x))$ has the properties discussed above in this section and, added to (\ref{eq:Mod_flow_chargechiralFF}), provides the modular flow of the total charge density $j(x)$, as given in (\ref{eq:modflow_totaldensity}).

We conclude this section by discussing the modular correlation functions. 
 The modular correlation function of a chiral field $\psi_+^\dagger$ and the one of the charge density $j_+$ are given by (\ref{eq:mod corrfunc FF_kernel}) and (\ref{FullmodularcorrelationFF_v2}) respectively, where the restricted correlator is now given by (\ref{eq:2p_corr_chiral}).
Starting from (\ref{eq:mod corrfunc FF_kernel}) and using the resolvent method, in \cite{Erdmenger:2020nop} the expression of
$G^{\textrm{\tiny (f)}}_{\textrm{\tiny mod}}$ has been found to be
\begin{equation}
\label{eq:MCF_FFchiral}
    G^{\textrm{\tiny (f,+)}}_{\textrm{\tiny mod}}(x,y;t)=
G^{\textrm{\tiny (+)}}_V(x,y)\frac{\sinh[\pi \tilde{t}(x,y)]}{\sinh[\pi(\tilde{t}(x,y)+t-\mathrm{i}0^+)]}\,,
\end{equation}
 consistently with the previous results in the literature \cite{Longo:2009mn,Hollands:2019hje}. From (\ref{FullmodularcorrelationFF_v2}), $G^{\textrm{\tiny (c,+)}}_{\textrm{\tiny mod}}$ defined in (\ref{eq:modcorr_chiraldensity}) is given by the square of (\ref{eq:MCF_FFchiral}) \cite{Hollands:2019hje, Mintchev:2020jhc}. 
 According to the discussion above, we can obtain $G^{\textrm{\tiny (c,$-$)}}_{\textrm{\tiny mod}}$ from $G^{\textrm{\tiny (c,+)}}_{\textrm{\tiny mod}}$ by replacing $t\to -t$.
From (\ref{eq:modcorr_totaldensity}), the sum of $G^{\textrm{\tiny (c,+)}}_{\textrm{\tiny mod}}$ and $G^{\textrm{\tiny (c,$-$)}}_{\textrm{\tiny mod}}$ gives the modular correlation function of $j(x)=j_+(x)+j_-(x)$.

\subsection{Symmetry-resolved modular correlation function through bosonization}
\label{subsec:SRmdocorrfunc_FF}

Consider the conserved $U(1)$ charge $Q$ defined in (\ref{eq:def_total charge}). Given that the ground state of the Dirac field theory is an eigenstate of $Q$ and the total charge density $j$ in (\ref{eq:def_total charge}) commutes with $Q$, it is possible to define a symmetry resolution for the modular correlation function of $j$. The computation of the symmetry-resolved modular correlation function of $j$ is the goal of this section. As in the previous part of this section, also in what follows we consider the operators appearing in the correlation function to be evaluated at the same time co-ordinate.
  
Exploiting (\ref{eq:SR_modcorr FF}), (\ref{eq:resum_SRMCF_FF}), (\ref{eq:def_total charge}), (\ref{eq:factor_modflow}) and the definition of $j(x)$ given above (\ref{eq:modflow_totaldensity}), we can decompose the modular correlation function in (\ref{eq:modcorr_totaldensity}) as
\bea
G^{\textrm{\tiny D}}_{\textrm{\tiny mod}}(t)
&=&
\sum_{q\in\mathbb{Z}}p_V(q)\left[
G^{\textrm{\tiny (1)}}_{\textrm{\tiny mod}}(x,y;q,t)
+
G^{\textrm{\tiny (2)}}_{\textrm{\tiny mod}}(x,y;q,t)
+G^{\textrm{\tiny (3)}}_{\textrm{\tiny mod}}(x,y;q,t)
+
G^{\textrm{\tiny (4)}}_{\textrm{\tiny mod}}
(x,y;q,t)
\right]
\nonumber
\\
&\equiv&
\sum_{q\in\mathbb{Z}}p_V(q)
G^{\textrm{\tiny D}}_{\textrm{\tiny mod}}(q,t)
\,,
\label{eq:fourcontributions}
\eea
where
\be
G^{\textrm{\tiny (1)}}_{\textrm{\tiny mod}}(x,y;q,t)
\equiv
\frac{1}{p_V(q)}
\int_{-\pi}^{\pi}\frac{d\alpha}{2\pi} 
e^{-\mathrm{i}\alpha q}
\langle\Omega|e^{\mathrm{i}\alpha \int_V j_+(x')
\mathrm{d}x'}
e^{\mathrm{i}\alpha \int_{V} j_-(y')
\mathrm{d}y'}
j_+(x)\sigma_t(j_+(y))|\Omega\rangle\,,
\label{G1_2dDirac}
\ee
\be
\label{G2_2dDirac}
G^{\textrm{\tiny (2)}}_{\textrm{\tiny mod}}(x,y;q,t)
\equiv
\frac{1}{p_V(q)}
\int_{-\pi}^{\pi}\frac{d\alpha}{2\pi} 
e^{-\mathrm{i}\alpha q}
\langle\Omega|e^{\mathrm{i}\alpha \int_V j_+(x')
\mathrm{d}x'}
e^{\mathrm{i}\alpha \int_{V} j_-(y')
\mathrm{d}y'}
j_+(x)
\bar{\sigma}_t(j_-(y))|\Omega\rangle\,,
\ee
\be
\label{G3_2dDirac}
G^{\textrm{\tiny (3)}}_{\textrm{\tiny mod}}(x,y;q,t)
\equiv
\frac{1}{p_V(q)}
\int_{-\pi}^{\pi}\frac{d\alpha}{2\pi} 
e^{-\mathrm{i}\alpha q}
\langle\Omega|e^{\mathrm{i}\alpha \int_V j_+(x')
\mathrm{d}x'}
e^{\mathrm{i}\alpha \int_{V} j_-(y')
\mathrm{d}y'}
\sigma_t(j_+(y))
j_-(x)|\Omega\rangle\,,
\ee
\be
\label{G4_2dDirac}
G^{\textrm{\tiny (4)}}_{\textrm{\tiny mod}}(x,y;q,t)
\equiv
\frac{1}{p_V(q)}
\int_{-\pi}^{\pi}\frac{d\alpha}{2\pi} 
e^{-\mathrm{i}\alpha q}
\langle\Omega|e^{\mathrm{i}\alpha \int_V j_+(x')
\mathrm{d}x'}
e^{\mathrm{i}\alpha \int_{V} j_-(y')
\mathrm{d}y'}j_-(x)\bar{\sigma}_t(j_-(y))|\Omega\rangle\,,
\ee
and, from (\ref{eq:prob_FT}) and (\ref{eq:def_total charge}),
\be
\label{eq:pq_FT2}
p_V(q)=
\int_{-\pi}^{\pi}\frac{d\alpha}{2\pi} 
e^{-\mathrm{i}\alpha q}
\langle\Omega|e^{\mathrm{i}\alpha \int_V j_+(x')
\mathrm{d}x'}
e^{\mathrm{i}\alpha \int_{V} j_-(y')
\mathrm{d}y'}|\Omega\rangle\,.
\ee
We can conclude that $G^{\textrm{\tiny D}}_{\textrm{\tiny mod}}(q,t)$  provides the symmetry-resolved modular correlation function of the total charge density $j$. 
A remarkable property we can observe from (\ref{eq:fourcontributions}) is that there are non-vanishing terms involving fields with different chiralities. This is different from what we found for the unresolved modular correlation function (\ref{eq:modcorr_totaldensity}).
Before computing the four terms in (\ref{eq:fourcontributions}), let us discuss the expression for the probability distribution $p_V(q)$, which reads
\be
\label{eq:pq_Dirac}
p_V(q)
=
\sqrt{\frac{\pi}{2 \ln L_p}}e^{-\frac{\pi^2 q^2}{2 \ln L_p}}
\,,
\ee
where
\be
\label{eq:Lp_def}
L_p\equiv\frac{\mathrm{\epsilon}^{-p}\prod_{i,j=1}^p |b_i-a_j|}{\prod_{i< j=1}^p|a_i-a_j||b_i-b_j|}
\,.
\ee
The Gaussian probability distribution in (\ref{eq:pq_Dirac}) has first been found in \cite{Murciano:2021djk,Foligno:2022ltu} for the two-dimensional free massless Dirac theory and its derivation is sketched in Appendix \ref{apx:derivations}.
Notice that, as the UV cutoff $\epsilon\to 0^+$, the dimensionless ratio $L_p$ in (\ref{eq:Lp_def}) diverges. In the forthcoming analysis we often expand the results in the parameter $L_p\gg 1$.

We now consider $G^{\textrm{\tiny (1)}}_{\textrm{\tiny mod}}(x,y;q,t)$ in (\ref{G1_2dDirac}) and, using the expression (\ref{eq:ModFlowSimple_4}) for the charge density $j_+$ in (\ref{eq:chiral charge densities}), we rewrite it as 
\be
\label{eq:G1_resolvent}
G^{\textrm{\tiny (1)}}_{\textrm{\tiny mod}}(x,y;q,t)
=
\frac{1}{p_V(q)}
\int_V
\mathrm{d}x_1\mathrm{d} x_2
\Sigma_t (x_1,y)\Sigma_t (x_2,y)
\int_{-\pi}^{\pi}\frac{d\alpha}{2\pi} 
e^{-\mathrm{i}\alpha q}
F^{\textrm{\tiny (1)}}(x,x_1,x_2;\alpha)\,,
\ee
where
\be
\label{eq:F1alpha}
F^{\textrm{\tiny (1)}}(x,x_1,x_2;\alpha)
\equiv
\langle\Omega|e^{\mathrm{i}\alpha \int_V j_+(x')
\mathrm{d}x'}
e^{\mathrm{i}\alpha \int_{V} j_-(y')
\mathrm{d}y'}
j_+(x):\psi_+^\dagger(x_1)\psi_+(x_2):|\Omega\rangle
\,.
\ee
The correlation function in (\ref{eq:F1alpha}) can be computed by resorting to bosonization techniques \cite{Tsvelikbook,DiFrancescobook}. Within this duality,
a free massless chiral fermion is mapped into a free massless chiral compact boson by
\be
\label{eq:bosonisationfermion}
\psi_+ \leftrightarrow \sqrt{\frac{1}{2\pi \epsilon}} \colon e^{-\mathrm{i}\phi_+}\colon
\equiv 
\sqrt{\frac{1}{2\pi \epsilon}}\mathcal{V}^{\textrm{\tiny (+)}}_{- 1}\,,
\qquad
\psi_- \leftrightarrow \sqrt{\frac{1}{2\pi \epsilon}} \colon e^{\mathrm{i}\phi_-}\colon
\equiv 
\sqrt{\frac{1}{2\pi \epsilon}}\mathcal{V}^{\textrm{\tiny $(-)$}}_{1}\,,
\ee
where $\epsilon$ is the UV cutoff and we have introduced the notation $\mathcal{V}_{a}^{\textrm{\tiny $(+)$}}$ and $\mathcal{V}^{\textrm{\tiny $(-)$}}_a$ for the chiral vertex operators. As a consequence of (\ref{eq:bosonisationfermion}), the charge densities of the chiral fermions defined in (\ref{eq:chiral charge densities}) can be written as
\be
\label{eq:bosonisationcharge}
j_+\leftrightarrow \frac{1}{2\pi}\partial \phi_+\,,
\qquad\qquad
j_-\leftrightarrow \frac{1}{2\pi}\partial \phi_-
\,.
\ee
Let us employ (\ref{eq:bosonisationcharge}) to rewrite the exponential operator in $F^{\textrm{\tiny (1)}}(x,x_1,x_2;\alpha)$ as \cite{Goldstein:2017bua}
\be
\label{eq:bosonisationFlux}
e^{\mathrm{i}\alpha \int_V j_+(x')
\mathrm{d}x'}
e^{\mathrm{i}\alpha \int_{V}j_-(y')
\mathrm{d}y'}
=
\prod_{j=1}^p 
\mathcal{V}^{\textrm{\tiny $(+)$}}_{\frac{\alpha}{2\pi}}( b_j)
\mathcal{V}^{\textrm{\tiny $(-)$}}_{\frac{\alpha}{2\pi}}(b_j)
\mathcal{V}^{\textrm{\tiny $(+)$}}_{-\frac{\alpha}{2\pi}}(a_j)
\mathcal{V}^{\textrm{\tiny $(-)$}}_{-\frac{\alpha}{2\pi}}(a_j)\,,
\ee
where we have made explicit use of the fact that $V$ is made by $p$ disjoint intervals, namely $V=\bigcup_{j=1}^p[a_j,b_j]$.
Using (\ref{eq:bosonisationfermion}), (\ref{eq:bosonisationcharge}) and (\ref{eq:bosonisationFlux}), the correlation function $F^{\textrm{\tiny (1)}}(x,x_1,x_2;\alpha)$ can be written in the following bosonic form
\be
\label{eq:Falpha_bosonised}
F^{\textrm{\tiny (1)}}(x,x_1,x_2;\alpha)
=\frac{
\left\langle\Omega\left|
\mathcal{V}^{\textrm{\tiny $(+)$}}_{\frac{\alpha}{2\pi}}( b_j)
\mathcal{V}^{\textrm{\tiny $(-)$}}_{\frac{\alpha}{2\pi}}(b_j)
\mathcal{V}^{\textrm{\tiny $(+)$}}_{-\frac{\alpha}{2\pi}}(a_j)
\mathcal{V}^{\textrm{\tiny $(-)$}}_{-\frac{\alpha}{2\pi}}(a_j)
\partial \phi_+(x) \colon \mathcal{V}^{\textrm{\tiny $(+)$}}_{1}(x_1) \mathcal{V}^{\textrm{\tiny $(+)$}}_{-1}(x_2) \colon
\right|\Omega\right\rangle}
{(2\pi)^2 \epsilon}\,.
\ee
The correlation functions of vertex and derivative operators are known and are reported in Appendix \ref{apx:derivations}. 
Exploiting the Wick's theorem and using (\ref{eq:derivative-scalar}), (\ref{eq:vertexoperator}) and (\ref{eq:vertexoperatorantichiral}), we obtain the following expression for (\ref{eq:F1alpha})
\be
\label{eq:F1alpha_v2}
F^{\textrm{\tiny (1)}}(x,x_1,x_2;\alpha)=
\frac{1}{(2\pi)^2 }\frac{H(x,x_1,x_2)+\frac{\alpha}{2\pi}J_p(x)}{(x_1-x_2-\mathrm{i\epsilon})} e^{-\frac{\alpha^2}{2\pi^2}\ln L_p-\frac{\alpha}{2\pi}K_p(x_1,x_2)}
\,,
\ee
where
\bea
\label{eq:Hdef}
H(x,x_1,x_2)&\equiv&\frac{x_2-x_1}{(x-x_1-\mathrm{i\epsilon})(x-x_2-\mathrm{i\epsilon})}\,,
\\
\label{eq:Jdef}
J_p(x)&\equiv& \sum_{j=1}^p
\left(
\frac{b_j-a_j}{(b_j-x-\mathrm{i\epsilon})(x-a_j+\mathrm{i\epsilon})}
\right)\,,
\\
\label{eq:Kdef}
K_p(x_1,x_2)&\equiv& \sum_{j=1}^p
\log \left[
\frac{ (b_j-x_2-\mathrm{i\epsilon})(a_j-x_1-\mathrm{i\epsilon})}{(b_j-x_1-\mathrm{i\epsilon})(a_j-x_2-\mathrm{i\epsilon})}
\right]\,,
\eea
and $L_p$ is defined in (\ref{eq:Lp_def}).
When $\alpha=0$, we obtain
\be
\label{eq:check alpha 0}
F^{\textrm{\tiny (1)}}(x,x_1,x_2;0)
=
-\frac{1}{(2\pi)^2} \frac{1}{(x-x_1-\mathrm{i\epsilon})(x-x_2-\mathrm{i\epsilon})}\,,
\ee
which is consistent with (\ref{Wickusing_v0}) when the two-point function (\ref{eq:2p_corr_chiral}) is employed.

In order to obtain $G^{\textrm{\tiny (1)}}_{\textrm{\tiny mod}}(x,y;q,t) $, we first have to compute the Fourier transform of (\ref{eq:F1alpha_v2}). We evaluate it in the limit $L_p\to\infty$,
within a saddle point approximation around the point $\alpha=0$. This allows to stretch the integration domain between $-\infty$ and $\infty$ 
and therefore the Fourier transform of (\ref{eq:F1alpha_v2}) is achieved through a Gaussian integral. The leading orders in the expansion for $L_p\to\infty$ read
\bea
\label{eq:FTF1leadings}
&&
\int_{-\pi}^{\pi}
\frac{d\alpha}{2\pi} 
e^{-\mathrm{i}\alpha q}
F^{\textrm{\tiny (1)}}(x,x_1,x_2;\alpha)
\\
&&=
\frac{p_V(q)}
{4\pi^2(x_1-x_2-\mathrm{i}\epsilon)}
\bigg[
H(x,x_1,x_2)
+
\frac{\mathrm{i}\pi q}{2\ln L_p}
\big(
H(x,x_1,x_2)K_p(x_1,x_2)-J_p(x)
\big)
\nonumber
\\
&&\hspace{0.4cm}+
\frac{1}{8\ln L_p}K_p(x_1,x_2)
\big(H(x,x_1,x_2)K_p(x_1,x_2)
-2 J_p(x)
\big)
+
O\big[(\ln L_p)^{-2}\big]
\bigg]
\,,
\nonumber
\eea
where $p_V(q)$ is given in (\ref{eq:pq_Dirac}) and $H$, $J_p$ and $K_p$ are defined in (\ref{eq:Hdef}), (\ref{eq:Jdef}) and (\ref{eq:Kdef}) respectively.
At this point (\ref{eq:FTF1leadings}) must be plugged back into (\ref{eq:G1_resolvent}) and the two integrals in the co-ordinates $x_1$ and $x_2$ must be performed. 
This complicated task simplifies when we restrict our interest to the leading term in the expansion for $L_p\gg 1$. We obtain
\be
\label{eq:G1leading}
G^{\textrm{\tiny (1)}}_{\textrm{\tiny mod}}(x,y;q,t)
=
\int_V
\mathrm{d}x_1\mathrm{d} x_2
\Sigma_t (x_1,y)\Sigma_t (x_2,y)G^{\textrm{\tiny $(+)$}}_V(x,x_1)G^{\textrm{\tiny $(+)$}}_V(x,x_2)
+
O\big[(\ln L_p)^{-1}\big]
\,.
\ee
Notice that the subleading terms vanish as the UV cutoff $\epsilon\to 0$. Nevertheless, their analysis for values of $\epsilon$ different from zero is interesting and deserves future investigation. 
The result in (\ref{eq:G1leading}) has various insightful features.
First, it turns out to be independent of the charge sector we are restricting to and the dependence on the charge arises in the corrections which vanish as $L_p\to\infty$. We will come back to this observation later on in the section. Second, 
along the analysis reported in Sec.\,\ref{subsec:modflowandcorrelator-FF_noSR} (see (\ref{eq:modcorr_chiraldensity}) and following lines), comparing (\ref{eq:G1leading}) with (\ref{FullmodularcorrelationFF_v2}), it is straightforward to realize that
\be
\label{eq:G1_Dirac_final}
G^{\textrm{\tiny (1)}}_{\textrm{\tiny mod}}(x,y;q,t)
= \left[G^{\textrm{\tiny (f,$+$)}}_{\textrm{\tiny mod}}(x,y;t)\right]^2+O\big[(\ln L_p)^{-1}\big]
=
 G^{\textrm{\tiny (c,$+$)}}_{\textrm{\tiny mod}}(x,y;t)+O\big[(\ln L_p)^{-1}\big]\,,
\ee
where $ G^{\textrm{\tiny (f,$+$)}}_{\textrm{\tiny mod}}$ is given in (\ref{eq:MCF_FFchiral}) and $ G^{\textrm{\tiny (c,$+$)}}_{\textrm{\tiny mod}}$ is defined in (\ref{eq:modcorr_chiraldensity}).

The other contributions $G^{\textrm{\tiny $(k)$}}_{\textrm{\tiny mod}}$ with $k=2,3,4$ (see (\ref{G2_2dDirac})-(\ref{G4_2dDirac})) can be computed with very similar steps. Here we report only the results, while in Appendix \ref{apx:derivations} an example of modular correlation function involving both the chiral fields $\psi_+$ and $\psi_-$, i.e. $G^{\textrm{\tiny (3)}}_{\textrm{\tiny mod}}$ in (\ref{G3_2dDirac}), is discussed more in detail.
We find
\bea
G^{\textrm{\tiny (2)}}_{\textrm{\tiny mod}}(x,y;q,t)
&=&
O\big[(\ln L_p)^{-1}\big]
\label{eq:G2_Dirac_final}
\,,
\\
G^{\textrm{\tiny (3)}}_{\textrm{\tiny mod}}(x,y;q,t)
&=&
O\big[(\ln L_p)^{-1}\big]
\,,
\label{eq:G3_Dirac_final}
\\
G^{\textrm{\tiny (4)}}_{\textrm{\tiny mod}}(x,y;q,t)
\label{eq:G4_Dirac_final}
&=&
G^{\textrm{\tiny (c,$-$)}}_{\textrm{\tiny mod}}(x,y,t)
+
O\big[(\ln L_p)^{-1}\big]\,,
\eea
where the definition of  $G^{\textrm{\tiny (c,$-$)}}_{\textrm{\tiny mod}}$ is given in (\ref{eq:modcorr_chiraldensity}).
Notice that the terms of $G^{\textrm{\tiny D}}_{\textrm{\tiny mod}}(q,t)$ that involve both chiral fermions, namely $G^{\textrm{\tiny (2)}}_{\textrm{\tiny mod}}$ and $G^{\textrm{\tiny (3)}}_{\textrm{\tiny mod}}$ in (\ref{G2_2dDirac}) and (\ref{G3_2dDirac}) respectively, are non-vanishing, but contribute at subleading order compared to $G^{\textrm{\tiny (1)}}_{\textrm{\tiny mod}}$ and $G^{\textrm{\tiny (4)}}_{\textrm{\tiny mod}}$. Given that such subleading terms go to zero when $L_p\to\infty$, they do not affect the following analysis. The leading contribution in (\ref{eq:G4_Dirac_final}), as already seen for the one in (\ref{eq:G1_Dirac_final}), is independent of the charge $q$, differently from the subleading orders.

Adding up (\ref{eq:G1_Dirac_final}), (\ref{eq:G2_Dirac_final}), (\ref{eq:G3_Dirac_final}) and (\ref{eq:G4_Dirac_final}), we obtain the modular correlation function (\ref{eq:modcorr_totaldensity}) resolved in the charge sector $q$, which reads
\be
\label{eq:SRmodcorrfunction_finres}
G^{\textrm{\tiny D}}_{\textrm{\tiny mod}}(q,t)
=
G^{\textrm{\tiny (c,$+$)}}_{\textrm{\tiny mod}}(x,y,t)
+
G^{\textrm{\tiny (c,$-$)}}_{\textrm{\tiny mod}}(x,y,t)
+
O\big[(\ln L_p)^{-1}\big]
=
G^{\textrm{\tiny D}}_{\textrm{\tiny mod}}
(t)
+
O\big[(\ln L_p)^{-1}\big]\,.
\ee
This is one of main result contained in this manuscript and therefore some comments are in order.
We first notice that the symmetry-resolved modular correlation function $G^{\textrm{\tiny D}}_{\textrm{\tiny mod}}(q,t)$ is equal to the unresolved one at leading order in the expansion for $\epsilon\to 0$, i.e. $L_p\to \infty$. This implies that, plugging back (\ref{eq:SRmodcorrfunction_finres}) into the right-hand side of (\ref{eq:fourcontributions}), the left-hand side is straightforwardly retrieved at leading order because of the normalization of the probability $p_V(q)$.
 Moreover, as already pointed out for the various contributions to $G^{\textrm{\tiny D}}_{\textrm{\tiny mod}}(q,t)$, the leading order is independent of the charge sector and the dependence on $q$ only arises at the subleading orders that are vanishing as $\epsilon\to 0$.
It is then natural to make connection with the so-called equipartition of the entanglement, which states that all the symmetry-resolved entanglement entropies are, at leading order in the UV cutoff, indpendent of the charge sector \cite{Xavier:2018kqb}. In this sense, we can say that (\ref{eq:SRmodcorrfunction_finres}) shows the {\it equipartition of the modular correlation function} of the charge density in the various symmetry sectors. It would be interesting to understand whether this result holds for more general modular correlation functions and more general theories or it is peculiar of the charge density in massless Dirac theories in $1+1$ dimensions. We leave this problem for future investigations.

\section{Conclusions}
\label{sec:Conclusion}

In this manuscript we study a possible resolution into $U(1)$ charge sectors of the modular flow and the corresponding modular correlation functions of $U(1)$-invariant operators. 
To the best of our knowledge, this is the first attempt in achieving the symmetry resolution for these quantities.
We consider a theory with a $U(1)$ global symmetry and a spatial bipartition into $V$ and its complement by associating algebras of operators to the two subregions. We assume that the Hilbert space can be factorized into two parts, associated to $V$ and its complement. This happens in systems defined in terms of finite dimensional algebras or, understood in a formal way to access well-defined results, in free QFTs. In this factorized Hilbert space, we consider states with a fixed value of the $U(1)$ charge generating the symmetry; this amounts to restrict our analysis to one of the terms occurring in the decomposition (\ref{eq:decHS}).
In the context of the Tomita-Takesaki theory, we achieve a modular relation of the form (\ref{eq:modularrelationfixedq}) for each value $q$ of the charge in the subregion $V$. 
Following the usual modular theory, this allows to define the modular operator (\ref{eq:SRmodop_new}) in each charge sector. 
In general, a modular operator is uniquely identified by a local algebra and a cyclic and separating vector.
In our analysis we provide the algebra as a suitable subalgebra of $U(1)$-invariant operators in $V$ and the state by projecting onto the Hilbert space with a fixed value of the charge in $V$.
The relation (\ref{eq:modularrelationfixedq}) leads to the symmetry-resolved modular flow (\ref{eq:SRmodflow_new}) and the symmetry-resolved modular correlation function (\ref{eq:SRmodularcorrelations}). Consistently with a proper notion of symmetry resolution, both of them must yield, once properly summed over all the charged sectors, the corresponding unresolved quantities. This happens in (\ref{eq:modflow_decomposition}) for the modular flow and in (\ref{eq:SRmodularcorrelations}) for the modular correlation function. Moreover, the symmetry-resolved modular correlation function satisfies the KMS condition (\ref{eq:SR_KMS condition}) for any value of the charge, as expected for a full-fledged modular correlation function.

As a playground to investigate the features of this symmetry resolution, we consider fermionic Gaussian states, where the modular flow and the modular correlation functions have been investigated in the literature for various settings and bipartitions. For this purpose, we adapt the approach developed in \cite{Erdmenger:2020nop} to the computation of the symmetry-resolved modular correlation functions.
 Since the symmetry resolution can be achieved only for the modular flow of $U(1)$-invariant operators, this cannot be accessed for a single fermionic field. The simplest starting point is the charge density operator for which we retrieve in (\ref{FullmodularcorrelationFF_v2}) the known results of \cite{Hollands:2019hje,Mintchev:2020jhc} on the modular correlation function through a different computation based on \cite{Erdmenger:2020nop}. This also allows a first simple application of the approach of \cite{Erdmenger:2020nop} to modular flows of composite operators.
Turning to symmetry resolution, in (\ref{eq:SRmodcorrfunc_FT}) we obtain an explicit formula for computing the symmetry-resolved modular correlation function of the charge density. This involves correlation functions of fermionic fields where neither the modular flow nor the modular Hamiltonian appear and that can be computed case by case.
As an explicit example, we consider a free massless Dirac theory in $1+1$-dimensions. Given the theory on an infinite line bipartite into a region made up of $p$ disjoint intervals and its complement, we compute the symmetry-resolved  modular correlation function of the charge density. The result is reported in (\ref{eq:SRmodcorrfunction_finres}) and exhibits a very interesting feature. Indeed, the leading order of (\ref{eq:SRmodcorrfunction_finres}) in the UV cutoff expansion, which is the only non-vanishing contribution as the cutoff is sent to zero, is independent of the charge sector, showing an equipartition analogous to the one observed in the context of symmetry-resolved entanglement \cite{Xavier:2018kqb}. A consequence of this feature is that all the properties valid for the unresolved modular correlator (such as the ones discussed in Sec.\,\ref{subsec:modflowandcorrelator-FF_noSR}) hold also for the symmetry-resolved one.

The derivations of the results of this manuscript rely on the definitions of reduced density matrices and charge operators restricted to spatial subregions, which, strictly speaking, are not allowed in the algebraic approach to QFT. Nonetheless, these objects can still be understood if we think of a UV-regularized theory, as we have done in Sec.\,\ref{sec:ChiralFF2d}. In order for the results to be well-defined, we have to check that their final expressions are meaningful when the UV cutoff is taken to zero. This is indeed what happens for the symmetry-resolved modular correlation function in (\ref{eq:SRmodcorrfunction_finres}), which, as already commented, is equal to the unresolved one that is a well-defined quantity in QFT. This indicates that the symmetry-resolved modular correlation function defined in Sec.\,\ref{subsec:AQFT_srmodflow} assuming finite-dimensional algebras may be well-defined also for Type III von Neumann algebras, which are necessary to define QFTs. To reinforce this statement, a derivation of the symmetry resolution that works directly for Type III algebras would be desirable.
A promising gradual approach towards this goal could be trying first to generalize the results of Sec.\,\ref{subsec:AQFT_srmodflow} to Type I$_\infty$ and Type II algebras.
We leave this interesting aspect for future investigations.

Our analysis inspires various questions that are worth to be addressed in follow-up works. One of the main findings of this manuscript is the equipartition of the symmetry-resolved modular correlation function of the charge density, obtained in a $1+1$-dimensional free massless Dirac theory. It would be interesting to investigate the robustness of this result by changing the $U(1)$-invariant operator evolved through the modular flow or the theory and the symmetry group under consideration.
Another intriguing extension of this work might regard the connection between the symmetry resolution and the relative modular operator. Starting from the latter, it is possible to define the so-called Connes-Radon-Nikodym (CRN) flow, which generalizes the modular flow. Differently from the modular flow, the definition of the CRN flow requires two states and has a tight connection with the relative entropy of such states \cite{Araki:1976zv}.
A preliminary analysis on the symmetry resolution of the CRN flow is reported in Appendix \ref{apx:relmodflow}, but a toolkit to explicitly compute it in some cases, as for instance in free fermionic theories, is still missing.

In light of recent progress on the applications of modular flows to the AdS/CFT correspondence, we comment on possible implications of the results of this manuscript on charge decomposition for AdS/CFT.
A first development in this context will be to adapt our analysis to holographic CFTs. For these, the von Neumann operator algebra is of  Type III$_1$ in the strict large $N$ limit \cite{Jefferson:2018ksk,Leutheusser:2021qhd,Leutheusser:2021frk,Chandrasekaran:2022eqq}. This  excludes an application of our discussion of Sec.\,\ref{sec:AlgebraicModularTheory} above, unless generalized to a framework that does not require reduced density matrices to be defined. On the other hand, in \cite{Witten:2021unn} it is shown that when including  $1/N$ corrections, the aforementioned Type III$_1$ algebra becomes of Type II$_\infty$, which is  closer to the framework of Sec.\,\ref{sec:AlgebraicModularTheory}. Moreover, Type II$_1$ algebras were used to describe gravitational theories in de Sitter spacetimes \cite{Chandrasekaran:2022cip}.
When extending the results in Sec.\,\ref{subsec:AQFT_srmodflow} to holographic CFTs, it will be interesting to understand if a charge sector decomposition for boundary operators such as the ones in (\ref{eq:Adecomposition}) can be identified also for the corresponding bulk fields.
This can be potentially achieved taking into account the decomposition (\ref{eq:modflow_decomposition}) in the context of holographic bulk reconstruction
\cite{Hamilton:2006az,Kabat:2011rz}. Indeed, as clarified in \cite{Jafferis:2015del, Faulkner:2017vdd}, properly smearing the modular flow of operators in a given boundary subregion, it is possible to reconstruct bulk operators in the corresponding entanglement wedge.
Furthermore, the recently explored connection between Berry phases and entanglement both in QFT and for AdS/CFT \cite{Nogueira:2021ngh,Banerjee:2022jnv} may be investigated exploiting symmetry-resolved modular flows.


In \cite{May:2018tir,Chandrasekaran:2021tkb} an expression relating a family of quantum information quantities, including relative entropies and fidelities, to certain modular correlation functions was derived and exploited in a holographic context.
Studying the interplay between these formulae and the symmetry resolution of the modular correlation functions discussed in Sec.\,\ref{subsec:AQFT_srmodflow} may allow to retrieve known examples of resolution of quantum information quantities \cite{Capizzi:2021zga,Parez:2022sgc}, as well as to derive new ones.
To frame the symmetry-resolved modular correlation functions in a holographic context, given the formula (\ref{eq:modular corr_step1}), we may find helpful introducing an Aharonov-Bohm flux, which, for instance, is realized in an AdS$_3$ bulk by a Wilson line defect anchored at the endpoints of the boundary subregion (see \cite{Zhao:2020qmn,Weisenberger:2021eby,Zhao:2022wnp} for the application of this idea to compute the holographic symmetry-resolved entanglement entropy).

Finally, though free, fermionic theories as the ones discussed in Sec.\,\ref{sec:FreeFermions} and Sec.\,\ref{sec:ChiralFF2d} are relevant in some AdS/CFT settings. For instance, in \cite{Gaberdiel:2013vva,Gaberdiel:2014cha} the duality between a class of two-dimensional supersymmetric coset CFTs and Vasiliev higher spin theories in AdS$_3$ is considered and it is shown that, in the limit of large level of these cosets, free fermionic and bosonic fields arise in the boundary theory.
Given the knowledge of the dual gravity theory, approaching our analysis from the bulk perspective in these models is a suggestive idea for follow-up works.

\bigskip

\noindent {\bf Acknowledgements}

\noindent We are indebted to Diego Pontello for several deep insights on algebraic quantum field theory and valuable comments on the manuscript.
We also thank Pablo Basteiro, Pasquale Calabrese, Moritz Dorband, Animik Ghosh, Ren\'e Meyer, Sara Murciano, Ignacio Reyes and
Erik Tonni for useful discussions.
We acknowledge support by the Deutsche Forschungsgemeinschaft (DFG, German Research Foundation) under Germany's Excellence Strategy through the W\"urzburg-Dresden Cluster of Excellence on Complexity and Topology in Quantum Matter - ct.qmat (EXC 2147, project-id 390858490), as well as
through a German-Israeli Project Cooperation (DIP) grant ”Holography and the Swampland”.

\begin{appendix}

\section{Sketch of the proof that \texorpdfstring{$|\Omega_q\rangle$}{Omegaq} is cyclic and separating}
\label{apx:omegaq_cyclic}

In this appendix we provide a sketch of the proof that the state $|\Omega_q\rangle$ defined in (\ref{eq:SandOmegaatfixedq_new}) is cyclic and separating for $\mathcal{A}_q(V)$. We stress that, crucially, the state $|\Omega\rangle$, to which the projectors are applied, is assumed to be cyclic and separating for $\mathcal{A}(V)$. 

Let us first us recall the definition of a cyclic and separating state. Given $|\Omega\rangle\in \mathcal{H}_{\bar{q}}$, where $\mathcal{H}_{\bar{q}}$ is defined through the decomposition (\ref{eq:decHS}), and $\mathcal{A}(V)\subset\mathcal{B}(\mathcal{H}_{\bar{q}})$, $|\Omega\rangle$ is cyclic for $\mathcal{A}(V)$ if the states $A|\Omega\rangle$, $A\in\mathcal{A}(V)$, are dense in $\mathcal{H}_{\bar{q}}$ . Moreover, it is separating for $\mathcal{A}(V)$ if $\nexists A\in\mathcal{A}(V)$ such that $A|\Omega\rangle=0$.
Now we consider $|\Omega_q\rangle$ given in (\ref{eq:SandOmegaatfixedq_new}) and we first prove that it is separating for $\mathcal{A}_q(V)$. Let us assume that an operator $A_q\otimes \boldsymbol{1}_{V'}\in \mathcal{A}_q(V)$ exists such that 
\be
\label{eq:absurdum}
A_q\otimes \boldsymbol{1}_{V'}|\Omega_q\rangle=0\,.
\ee
Using (\ref{eq:Adecomposition}), (\ref{eq:SandOmegaatfixedq_new}), (\ref{eq:projectorsonGS}) and the idempotence of the projectors, (\ref{eq:absurdum}) implies
\be
\label{eq:absurdum_2}
A_q\otimes \boldsymbol{1}_{V'}|\Omega\rangle=0
\,.
\ee
Since $\mathcal{A}_q(V)\subset \mathcal{A}(V)$, (\ref{eq:absurdum_2}) would imply that we have found an operator in $\mathcal{A}(V)$ which annihilates $|\Omega\rangle$. This is impossible since, by hypothesis, $|\Omega\rangle$ is separating for $\mathcal{A}(V)$ and therefore we conclude that $|\Omega_q\rangle$ must be separating for $\mathcal{A}_q(V)$.

In order to prove that $|\Omega_q\rangle$ is cyclic for $\mathcal{A}_q(V)$, we recall that the fact that $|\Omega\rangle$ is cyclic for $\mathcal{A}(V)$ implies $\mathcal{H}_{\bar{q}}=\overline{\mathcal{A}(V)|\Omega\rangle}$.
Exploiting (\ref{eq:decompositionalgebra}), (\ref{eq:SandOmegaatfixedq_new}) and (\ref{eq:projectorsonGS}), we obtain
\be
\mathcal{H}_{\bar{q}}=\sum_{q\in\mathbb{Z}}\overline{\mathcal{A}_q(V)|\Omega_q\rangle}\,.
\ee
Projecting $\mathcal{H}_{\bar{q}}$ onto $\mathcal{H}^{(q)}_{V}\otimes \mathcal{H}^{(\bar{q}-q)}_{V'}$ and using (\ref{eq:Adecomposition}), the idempotence and the orthogonality of the projectors, we find
\be
\mathcal{H}^{(q)}_{V}\otimes \mathcal{H}^{(\bar{q}-q)}_{V'}=\left[\Pi_V(q)\otimes \Pi_{V'}(\bar{q}-q)\right]\mathcal{H}_{\bar{q}}=
\overline{\mathcal{A}_q(V)|\Omega_q\rangle}
\,,
\ee
which implies that $|\Omega_q\rangle$ is cyclic for $\mathcal{A}_q(V)$.

\section{Relative modular operator and symmetry resolution}
\label{apx:relmodflow}

In this appendix, in the spirit of the resolution of the modular flow discussed in Sec.\,\ref{subsec:AQFT_srmodflow}, we examine the symmetry resolution for the Connes-Radon-Nikodym (CRN) flow of two states. We begin with a brief review of the relation between the CRN flow and the relative modular operator before discussing the results.

\subsection{Relative modular operator and Connes-Radon-Nikodym cocycle}

In this subsection we review the concepts of relative modular operator and CRN cocycle, both closely related to the relative entropy of a pair of states.
Consider a net of local algebras defining a QFT and one of its algebras $\mathcal{A}(V)\subset\mathcal{B}(\mathcal{H})$ associated to the causally complete spacetime region $V$. Consider also two states that can be represented by cyclic and separating vectors $|\Omega\rangle$ and $|\tilde{\Omega}\rangle$ in the Hilbert space $\mathcal{H}$.
Along the line of the modular theory summarized in Sec.\,\ref{subsec:TTtheory}, one can show that a unique antilinear {\it relative modular involution} exists such that \cite{Haagbook}
\be
\label{eq:relativemodular relation}
S_{\tilde{\Omega},\Omega} A |\Omega\rangle= A^\dagger|\tilde{\Omega}\rangle\,,
\qquad
\forall A\in \mathcal{A}(V)\,.
\ee
Through the polar decomposition $S_{\tilde{\Omega},\Omega}=J_{\tilde{\Omega},\Omega} \Delta^{1/2}_{\tilde{\Omega},\Omega} $, we can then introduce the {\it the relative modular conjugation} $J_{\tilde{\Omega},\Omega} $ and the {\it relative modular operator} $\Delta_{\tilde{\Omega},\Omega} $, where the latter is positive and self-adjoint.
The relative modular operator is a generalization of the modular operator, in the sense that it reduces to it when $|\Omega\rangle=|\tilde{\Omega}\rangle$.
We do not list here all the properties of the relative modular operator; a complete discussion can be found, for instance, in \cite{Haagbook}.

Given the modular operator $\Delta_{\Omega}$ associated to the state $|\Omega\rangle$,
the operator
\be
\label{eq:uoperator}
u_{\tilde{\Omega},\Omega}(t)
\equiv
\Delta_{\tilde{\Omega},\Omega}^{\textrm{i}t}\Delta^{-\textrm{i}t}_{\Omega}\,,
\ee
 is known as {\it CRN cocycle} \cite{Haagbook}. The CRN cocycle has been recently investigated also in the context of the AdS/CFT correspondence, where its gravity dual has been proposed \cite{Bousso:2020yxi}.  First, let us observe that $u_{\tilde{\Omega},\Omega}(t)\in \mathcal{A}(V)$ (notice that this does not happen for $\Delta_{\Omega}^{\textrm{i}t}$ and $\Delta_{\tilde{\Omega},\Omega}^{\textrm{i}t}$ separately).
Other important properties of the CRN cocycle are the following \cite{Haagbook}:
\begin{enumerate}
    \item Cocycle identity
    \be
    \label{eq:cocycle}
u_{\tilde{\Omega},\Omega}(t+s)=
u_{\tilde{\Omega},\Omega}(t)\,
\Delta^{\textrm{i}t}_{\Omega}u_{\tilde{\Omega},\Omega}(s)\Delta^{-\textrm{i}t}_{\Omega}\,,
\qquad\forall t,s\in \mathbb{R};
    \ee   
\item
Chain rule: given three cyclic and separating states $|\Omega\rangle$, $|\tilde{\Omega}\rangle$ and $|\hat{\Omega}\rangle$ in the Hilbert space $\mathcal{H}$, we have
\be
\label{eq:chainrule}
u_{\tilde{\Omega},\Omega}(t)=u_{\tilde{\Omega},\hat{\Omega}}(t)u_{\hat{\Omega},\Omega}(t)\,,
\qquad\forall t\in \mathbb{R};
\ee    
    \item
    Intertwining property:
    \be
    \label{eq:intertwinement}
    u_{\tilde{\Omega},\Omega}(t)\Delta^{\textrm{i}t}_{\Omega}A\Delta^{-\textrm{i}t}_{\Omega}
    =
\Delta^{\textrm{i}t}_{\tilde{\Omega}}A\Delta^{-\textrm{i}t}_{\tilde{\Omega}}
    u_{\tilde{\Omega},\Omega}(t)
    \,,
    \qquad\forall A\in \mathcal{A}(V),\;\;\forall t\in \mathbb{R}.
    \ee
\end{enumerate}
The last remarkable property of the CRN cocycle concerns its relation with the relative entropy, namely \cite{Araki:1976zv}
\be
\label{eq:relativeentropy_AQFT}
S_V(\Omega||\tilde{\Omega})\equiv
-\langle\Omega|\ln\Delta_{\tilde{\Omega},\Omega}|\Omega\rangle
=
\mathrm{i}\frac{d}{dt}\langle \Omega|u_{\tilde{\Omega},\Omega}(t)|\Omega\rangle|_{t=0}
\,.
\ee
Using the CRN cocycle, one can define a notion of CRN flow. Given an element $A$ of $\mathcal{A}(V)$, the CRN flow is defined as
\begin{equation}
\label{eq:trueCRNflow_AQFT}
u_{\tilde{\Omega},\Omega}(t)\Delta^{\textrm{i}t}_{\Omega}A\Delta^{-\textrm{i}t}_{\Omega}\,.
\end{equation}
Given that $\Delta^{\textrm{i}t}_{\Omega}A\Delta^{-\textrm{i}t}_{\Omega}$ and $u_{\tilde{\Omega},\Omega}(t)$ both belong to $\mathcal{A}(V)$, the CRN flow is an inner automorphism of $\mathcal{A}(V)$. 
If the factorization $\mathcal{H}=\mathcal{H}_V\otimes\mathcal{H}_{V'}$ holds, the relative modular operator and the CRN cocycle can be written in terms of the reduced density matrices $\rho_V$ and $\tilde{\rho}_V$ of the states $|\Omega\rangle$ and $|\tilde{\Omega}\rangle$ respectively. They read \cite{Haagbook}
\be
\label{eq:modopRDM}
\Delta_{\tilde{\Omega},\Omega}=
\tilde{\rho}_{V}\otimes\rho_{V'}^{-1}
\,,
\qquad\qquad
u_{\tilde{\Omega},\Omega}(t)=\tilde{\rho}_{V}^{\mathrm{i}t}\rho_{V}^{-\mathrm{i}t}\otimes \boldsymbol{1}_{V'}\,,
\ee
and, for any $A\otimes \boldsymbol{1}_{V'}\in\mathcal{A}(V)$, the CRN flow is
\be
\label{eq:relmodflowRDM}
\tilde{\rho}_{V}^{\mathrm{i}t}A\rho_{V}^{-\mathrm{i}t}\otimes \boldsymbol{1}_{V'}
\equiv
\sigma^{\textrm{\tiny $\tilde{\Omega},\Omega$}}_t(A)\otimes \boldsymbol{1}_{V'}
\,,
\ee
which reduces to (\ref{eq:modflowRDM}) when $|\Omega\rangle=|\tilde{\Omega}\rangle$.


\subsection{Symmetry resolution}

In order to define a notion of symmetry resolution for the CRN flow (\ref{eq:relmodflowRDM}), we consider the setup of Sec.\,\ref{subsec:AQFT_srmodflow}. The algebra $\mathcal{A}(V)$ is assumed to belong to the net of local observable algebras of a theory with a global $U(1)$ symmetry and therefore contains only $U(1)$-invariant operators.
Moreover, we also assume all the properties of the Hilbert space $\mathcal{H}$ discussed in Sec.\,\ref{sec:AlgebraicModularTheory}, including the factorization into $\mathcal{H}=\mathcal{H}_V\otimes\mathcal{H}_{V'}$ and the charge decompositions (\ref{eq:decHS}) and (\ref{eq:Hq_dec}).
We finally choose both the vectors involved in (\ref{eq:relativemodular relation}) to be eigenvectors of the generator of the $U(1)$ symmetry with eigenvalue $\bar{q}$, namely $|\Omega\rangle,|\tilde{\Omega}\rangle\in\mathcal{H}_{\bar{q}}$.

Following closely the derivation of (\ref{eq:modularrelationfixedq}), we check that, assuming (\ref{eq:relativemodular relation}), for each charge sector labeled by $q$ in the decompositions (\ref{eq:Hq_dec}) and (\ref{eq:Adecomposition}), we obtain
\be
 S_{\tilde{\Omega},\Omega,q} \left(A_q\otimes \boldsymbol{1}_{V'}\right) |\Omega_q\rangle= \left(A^\dagger_q\otimes \boldsymbol{1}_{V'}\right)|\tilde{\Omega}_q\rangle\,,
\ee
where
\begin{equation}
\label{eq:def_omegaq_relative}
S_{\tilde{\Omega},\Omega,q}=S_{\tilde{\Omega},\Omega}\left[\Pi_V(q)\otimes \Pi_{V'}(\bar{q}-q)\right]\,,
\end{equation}
\begin{equation}
  |\Omega_q\rangle=\left[\Pi_V(q)\otimes \Pi_{V'}(\bar{q}-q)\right]|\Omega\rangle\,,
  \qquad
|\tilde{\Omega}_q\rangle=\left[\Pi_V(q)\otimes \Pi_{V'}(\bar{q}-q)\right]|\tilde{\Omega}\rangle\,,
\end{equation}
and $\Pi_V(q)$ is given in (\ref{eq:FourierTransform}).
The operator in (\ref{eq:def_omegaq_relative}) can be decomposed as $S_{\tilde{\Omega},\Omega,q}=J_{\tilde{\Omega},\Omega,q}\Delta^{1/2}_{\tilde{\Omega},\Omega,q}$, where $\Delta_{\tilde{\Omega},\Omega,q}$ is connected to the relative modular operator by
  \begin{equation}
  \label{eq:srrelativmodop}
\Delta_{\tilde{\Omega},\Omega,q}=\Delta_{\tilde{\Omega},\Omega}\left[\Pi_V(q)\otimes \Pi_{V'}(\bar{q}-q)\right]\,.  
  \end{equation}
In our setup, the factorization $\mathcal{H}=\mathcal{H}_V\otimes\mathcal{H}_{V'}$ holds and therefore we can exploit (\ref{eq:modopRDM}) and (\ref{eq:relmodflowRDM}). Combining these equations with (\ref{eq:srrelativmodop}), we obtain
\be
\label{eq:modopRDM_sr}
\Delta_{\tilde{\Omega},\Omega,q}=
\frac{\tilde{p}_V(q)}{p_{V}(q)}\tilde{\rho}_{V}(q)\otimes\left[\rho_{V'}(\bar{q}-q)\right]^{-1}
\,,
\ee
where $p_V(q)$ is defined in (\ref{eq:SRRDM}) and $\tilde{p}_V(q)$ has the same definition but referred to the reduced density matrix $\tilde{\rho}_V$.
In analogy with (\ref{eq:modopRDM}), we define the CRN cocycle projected onto a given charge sector as
\be
\label{eq:srCRNcocycle}
u_{\tilde{\Omega},\Omega}(t,q)\equiv\left[\tilde{\rho}_{V}(q)\right]^{\mathrm{i}t}\left[\rho_{V}(q)\right]^{-\mathrm{i}t}\otimes \boldsymbol{1}_{V'}\,.
\ee
Considering the various terms in the decomposition (\ref{eq:Adecomposition}) of the operator $A\otimes \boldsymbol{1}_{V'}\in\mathcal{A}(V)$, we can define the {\it symmetry-resolved CRN flow} as
\be
\label{eq:relmodflowRDM_sr}
u_{\tilde{\Omega},\Omega}(t,q)
\Delta_{\Omega,q}^{\mathrm{i}t}\left(A_q\otimes\boldsymbol{1}_{V'}\right)\Delta_{\Omega,q}^{-\mathrm{i}t}
=\left[\tilde{\rho}_{V}(q)\right]^{\mathrm{i}t}A_q\left[\rho_{V}(q)\right]^{-\mathrm{i}t}\otimes \boldsymbol{1}_{V'}
\equiv\sigma^{\textrm{\tiny $\tilde{\Omega},\Omega$}}_{t,q}(A_q)\otimes \boldsymbol{1}_{V'}\,,
\ee
where in the first step we have exploited (\ref{eq:srCRNcocycle}) and (\ref{eq:SRmodop_new}).
Using (\ref{eq:SRRDM}) for both $\rho_V$ and $\tilde{\rho}_V$ in (\ref{eq:relmodflowRDM}) and (\ref{eq:relmodflowRDM_sr}), we find
\be
\label{eq:srrelmodflow}
\sigma^{\textrm{\tiny $\tilde{\Omega},\Omega$}}_t(A)\otimes \boldsymbol{1}_{V'}
=
\sum_{q\in\mathbb{Z}}
\left(\frac{\tilde{p}_V(q)}{p_{V}(q)}\right)^{\mathrm{i}t}
\sigma^{\textrm{\tiny $\tilde{\Omega},\Omega$}}_{t,q}(A_q)\otimes \boldsymbol{1}_{V'}
\,.
\ee
Notice that when $|\Omega\rangle=|\tilde{\Omega}\rangle$, (\ref{eq:relmodflowRDM_sr}) becomes (\ref{eq:SRmodflow_new}), (\ref{eq:srrelmodflow}) reduces to (\ref{eq:modflow_decomposition}) and we retrieve the results of Sec.\,\ref{subsec:AQFT_srmodflow}.

To conclude this appendix, we connect the findings reported here with other results from the literature.
As reported in (\ref{eq:relativeentropy_AQFT}), the relative entropy can be written in terms of $u_{\tilde{\Omega},\Omega}(t)$. By employing the CRN cocycle (\ref{eq:srCRNcocycle}) in the sector with charge $q$, we can define the symmetry-resolved relative entropy as
\begin{equation}
\label{eq:SRrelentropy_AQFT}
S_V(\rho_V(q)||\tilde{\rho}_V(q))
    \equiv\mathrm{i}\frac{d}{dt}
    \frac{\langle \Omega_q|u_{\tilde{\Omega},\Omega}(t,q)|\Omega_q\rangle}{\langle \Omega_q|\Omega_q\rangle}\bigg|_{t=0}
\,.
\end{equation}
Exploiting (\ref{eq:uoperator}), (\ref{eq:Adecomposition}), (\ref{eq:srrelativmodop}),  (\ref{eq:modopRDM_sr}), (\ref{eq:SRmodoperator}), (\ref{eq:SRmodop_new}) and (\ref{eq:normOmegaq}) in (\ref{eq:relativeentropy_AQFT}), after a bit of algebra, we find the decomposition
\begin{equation}
S_V(\rho_{V}||\tilde{\rho}_{V})=\sum_{q\in\mathbb{Z}} p_V(q)\left[S_V(\rho_{V}(q)||\tilde{\rho}_{V}(q))+\log\left(\frac{p_V(q)}{\tilde{p}_V(q)}\right)\right]
    \,,
\end{equation}
which is the one already found and studied in \cite{Capizzi:2021zga}.
In other words, the projection onto different charge sectors of the relative modular operator and CRN cocycle performed in this appendix is consistent with the definition of symmetry-resolved relative entropy existing in the literature.

\section{Gaussian states of \texorpdfstring{$N$}{N} non-interacting fermions}
\label{apx:NFreeFermions}

In this appendix we perform a straightforward generalisation of the results discussed in Sec.\,\ref{sec:FreeFermions} to the case of $N$ non-interacting fermions in a Gaussian state.
Let us collect the $N$ fermionic fields into a vector $\Psi=\left(\psi_1,\dots,\psi_N\right)^{\textrm{t}}$. Given that the fermions are non-interacting, 
we have the factorisation of the total density matrix and of any possible reduced density matrix  $\rho_V^{(\Psi)}$ of a given subsystem $V$, 
namely
\be
\label{RDM N fields}
\rho_V^{(\Psi)}=\bigotimes_{i=1}^N\rho_V^{(i)}\,,
\ee
where $\rho_V^{(i)}$ is the reduced density matrix associated to the single field $\psi_i$.
The charge density for this system is given by the composite operator
\be
\label{charge N fields}
\colon\Psi^\dagger(x) \Psi(x)\colon
=
\sum_{i=1}^N
\colon\psi_i^\dagger(x) \psi_i(x)\colon.
\ee
Using (\ref{charge N fields}) and (\ref{RDM N fields}), the modular flow of $\colon\Psi^\dagger(x) \Psi(x)\colon$, denoted by $\sigma^{\textrm{\tiny $(N)$}}_t$ to distinguish it from the modular flow of the charge density of a single fermionic field, is written as
\bea
&&
\sigma^{\textrm{\tiny $(N)$}}_t\left(\!\colon\Psi^\dagger(x) \Psi(x)\colon\!\right)
\equiv
\left[\rho_V^{(\Psi)}\right]^{\textrm{i}t}
\colon\Psi^\dagger(x) \Psi(x)\colon \left[\rho_V^{(\Psi)}\right]^{-\textrm{i}t}
\\
&&=
\sum_{j=1}^N
\left[\rho_V^{(j)}\right]^{\textrm{i}t}
\colon\psi_j^\dagger(x) \psi_j(x)\colon \left[\rho_V^{(j)}\right]^{-\textrm{i}t}
\equiv
\sum_{j=1}^N
\colon\sigma^{(j)}_t\left(\psi_j^\dagger(x) \psi_j(x)\right)\colon\,,
\eea
where we have also used that, as detailed in (\ref{eq:normal ordering}), for charge densities of single fermionic fields the normal ordering can be taken out from the the modular flow. 
If the fermionic fields are identical, the modular flow $\sigma^{(j)}_t\left(\psi_j^\dagger(x) \psi_j(x)\right)$ does not depend on the index $j$ and it is given by (\ref{eq:ModFlowSimple_4}). Thus, we can conclude that
\be
\sigma^{\textrm{\tiny $(N)$}}_t\left(\!\colon\Psi^\dagger(x) \Psi(x)\colon\!\right)
=
N\colon \sigma_t\left(\psi^\dagger(x) \psi(x)\right) \colon\,,
\ee
and therefore, up to a factor $N$, it enjoys all the properties discussed in Sec.\,\ref{sec:FreeFermions}.

Let us also comment the modular correlation function of the composite operator $\colon\Psi^\dagger(x) \Psi(x)\colon$. By using the definition (\ref{charge N fields}), we can write 
\bea
&&
\left\langle\Omega\big|\colon\Psi^\dagger(x) \Psi(x)\colon \sigma^{\textrm{\tiny $(N)$}}_t\left(\colon\Psi^\dagger(y) \Psi(y)\colon\right)\big|\Omega\right\rangle
\nonumber
\\
&&=
\sum_{i,j=1}^N
 \left\langle\Omega\big|\colon\psi_i^\dagger(x) \psi_i(x)\colon \colon\sigma^{(j)}_t\left(\psi_j^\dagger(y) \psi_j(y)\right)\colon\big|\Omega\right\rangle
 \nonumber
 \\
 &&=
 \sum_{i=1}^N
 \left\langle\Omega\big|\colon\psi_i^\dagger(x) \psi_i(x)\colon \colon\sigma^{(i)}_t\left(\psi_i^\dagger(y) \psi_i(y)\right)\colon\big|\Omega\right\rangle
 \nonumber
 \\
 &&\;\;\;\;\:+
 \sum_{i\neq j=1}^N
 \left\langle\Omega\big|\colon\psi_i^\dagger(x) \psi_i(x)\colon \colon\sigma^{(j)}_t\left(\psi_j^\dagger(y) \psi_j(y)\right)\colon\big|\Omega\right\rangle\,.
 \label{eq:MCF_2sums}
\eea
In the second sum, since $i\neq j$, because of (\ref{RDM N fields}) the correlators factorise into
\bea
 &&\left\langle\Omega\big|\colon\psi_i^\dagger(x) \psi_i(x)\colon \colon\sigma^{(j)}_t\left(\psi_j^\dagger(y) \psi_j(y)\right)\colon\big|\Omega\right\rangle
 \\
 &&=
 \nonumber
 \left\langle\Omega\big|\colon\psi_i^\dagger(x) \psi_i(x)\colon\big|\Omega\right\rangle\left\langle\Omega\big| \colon\sigma^{(j)}_t\left(\psi_j^\dagger(y) \psi_j(y)\right)\colon\big|\Omega\right\rangle
 =
 0\,,
\eea
and they are zero because of the normal ordering.
On the other hand, all the terms in the first sum in (\ref{eq:MCF_2sums}) are equal and independent of $i$ since the fermionic fields are identical. Thus, we finally
get
\be
\left\langle\Omega\big|\colon\Psi^\dagger(x) \Psi(x)\colon \sigma^{\textrm{\tiny $(N)$}}_t\left(\!\colon\Psi^\dagger(y) \Psi(y)\colon\!\right)\big|\Omega\right\rangle
=
N  \left\langle\Omega\big|\colon\psi^\dagger(x) \psi(x)\colon \colon\sigma_t\left(\psi^\dagger(y) \psi(y)\right)\colon\big|\Omega\right\rangle.
\ee
Summarising, in the case of $N$ non-interacting fermions in a Gaussian state both the modular flow of the charge density and its modular correlation function are simply the results for a single Gaussian fermion multiplied by $N$.
It is straightforward to see that the symmetry-resolved modular flow and modular correlation functions are the same, up to a factor $N$, as the ones
discussed in Sec.\,\ref{sec:FreeFermions} and Sec.\,\ref{sec:ChiralFF2d} for a single fermionic field.

\section{Computational details on the \texorpdfstring{$1+1$}{1+1}-dimensional massless chiral fermion}
\label{apx:derivations}

In this appendix we report some computational details of the derivation of the results contained in Sec.\,\ref{sec:ChiralFF2d}.

\subsection{Correlation functions}

In Sec.\,\ref{subsec:SRmdocorrfunc_FF} we have employed bosonization techniques to compute the symmetry-resolved modular correlation function of the charge density in a $1+1$-dimensional free massless Dirac field theory. In this first part of the appendix, we report some known correlation functions of the bosonic fields that we have exploited in our computation.

As in Sec.\,\ref{subsec:SRmdocorrfunc_FF}, we denote $\phi_+(x)$ and $\phi_-(x)$ the two chiral bosonic fields evaluated on the same constant time slice.
The correlation functions of a scalar field and its derivative are simply obtained starting from the two-point functions 
\cite{DiFrancescobook}
\be
\left\langle\Omega\left|
\phi_+(x)\phi_+(y)
\right|\Omega\right\rangle
-
\left\langle\Omega\left|
\phi_+(0)^2
\right|\Omega\right\rangle
=
-\log\left(\frac{x-y-\mathrm{i}\epsilon}{-\mathrm{i}\epsilon}\right)\,,
\ee
\be
\left\langle\Omega\left|
\phi_-(x)\phi_-(y)
\right|\Omega\right\rangle
-
\left\langle\Omega\left|
\phi_-(0)^2
\right|\Omega\right\rangle
=
-\log\left(\frac{x-y+\mathrm{i}\epsilon}{\mathrm{i}\epsilon}\right)\,,
\ee
and then taking the derivatives
\be
\label{eq:derivative-scalar}
\left\langle\Omega\left|
\partial\phi_+(x)\phi_+(y)
\right|\Omega\right\rangle
=
-\frac{1}{x-y-\mathrm{i}\epsilon}\,,
\quad\;
\left\langle\Omega\left|
\partial\phi_-(x)\phi_-(y)
\right|\Omega\right\rangle
=
-\frac{1}{x-y+\mathrm{i}\epsilon}\,.
\ee
Another useful result is the $n$-point function of vertex operators. For the vertex operators $\mathcal{V}^{\textrm{\tiny $(+)$}}_{a} $ and $\mathcal{V}^{\textrm{\tiny $(-)$}}_{a} $ considered in Sec.\,\ref{subsec:SRmdocorrfunc_FF}, we have \cite{DiFrancescobook}
\be
\label{eq:vertexoperator}
\left\langle\Omega\left|
 \mathcal{V}^{\textrm{\tiny $(+)$}}_{a_1}(x_1)\dots \mathcal{V}^{\textrm{\tiny $(+)$}}_{a_n}(x_n)
\right|\Omega\right\rangle
=
\delta_{0,\sum_{j=1}^n a_j}
\prod_{i<j=1}^n
\left(
\frac{x_i-x_j-\mathrm{i}\epsilon}{-\mathrm{i}\epsilon}
\right)^{a_i a_j}\,,
\ee
and
\be
\label{eq:vertexoperatorantichiral}
\left\langle\Omega\left|
 \mathcal{V}^{\textrm{\tiny $(-)$}}_{a_1}(x_1)\dots \mathcal{V}^{\textrm{\tiny $(-)$}}_{a_n}(x_n)
\right|\Omega\right\rangle
=
\delta_{0,\sum_{j=1}^n a_j}
\prod_{i<j=1}^n
\left(
\frac{x_i-x_j+\mathrm{i}\epsilon}{\mathrm{i}\epsilon}
\right)^{a_i a_j}\,.
\ee
The formulas (\ref{eq:derivative-scalar}), (\ref{eq:vertexoperator}) and (\ref{eq:vertexoperatorantichiral}) have been employed for computing (\ref{eq:Falpha_bosonised}) and the other correlation functions appearing in (\ref{G1_2dDirac})-(\ref{G4_2dDirac}).

In order to make the manuscript self-contained, let us use the correlation functions reported above to sketch the derivation of (\ref{eq:pq_Dirac}), namely the probability $p_V(q)$ of measuring charge $q$ in the subsystem made up of $p$ disjoint intervals in a $1+1$-dimensional massless Dirac field theory. 
We start from (\ref{eq:pq_FT2}) and we notice that the correlation function in the the integral can be rewritten through (\ref{eq:bosonisationFlux}) as
\bea
\langle\Omega|\prod_{j=1}^p \mathcal{V}^{\textrm{\tiny $(+)$}}_{\frac{\alpha}{2\pi}}(b_j)
\mathcal{V}^{\textrm{\tiny $(-)$}}_{\frac{\alpha}{2\pi}}(b_j)
\mathcal{V}^{\textrm{\tiny $(+)$}}_{-\frac{\alpha}{2\pi}}(a_j)
\mathcal{V}^{\textrm{\tiny $(-)$}}_{-\frac{\alpha}{2\pi}}(a_j)|\Omega\rangle
&=&
\left(\frac{\mathrm{\epsilon}^{-p}\prod_{i,j=1}^p |b_i-a_j|}{\prod_{i< j=1}^p|a_i-a_j||b_i-b_j|}\right)^{-\frac{\alpha^2}{2\pi^2}}
\nonumber
\\
&=&
\label{eq:correlationvertex_pq}
e^{-\frac{\alpha^2}{2\pi^2}\ln L_p}
\,,
\eea
where the right-hand side is obtained 
by exploiting (\ref{eq:vertexoperator}) and (\ref{eq:vertexoperatorantichiral}) and in the last step we have used (\ref{eq:Lp_def}). Plugging back (\ref{eq:correlationvertex_pq}) into (\ref{eq:pq_FT2}) and performing the Gaussian integral applying the saddle-point approximation around $\alpha=0$, we obtain (\ref{eq:pq_Dirac}),
which has been first obtained with similar techniques in \cite{Murciano:2021djk,Foligno:2022ltu}.

\subsection{Modular correlators involving different chiral fields }

In the final part of this appendix we report the main steps necessary for the computation of $G^{\textrm{\tiny (3)}}_{\textrm{\tiny mod}}(x,y;q,t)$ defined in (\ref{G3_2dDirac}). Differently from $G^{\textrm{\tiny (1)}}_{\textrm{\tiny mod}}(x,y;q,t)$ computed in detail in the main text, $G^{\textrm{\tiny (3)}}_{\textrm{\tiny mod}}(x,y;q,t)$ involves both the chiral fields $\psi_+$ and $\psi_-$. We recall that $G^{\textrm{\tiny (3)}}_{\textrm{\tiny mod}}(x,y;q,t)$, as the other contributions defined and computed in the main text, is defined for two operators evaluated at the same time co-ordinate.

Given that the ground state of the free massless Dirac theory is Gaussian, we can rely on the results discussed in Sec.\,\ref{sec:FreeFermions} for the modular correlation function of charge density operators.
Exploiting (\ref{eq:ModFlowSimple_4}) in (\ref{G3_2dDirac}), we obtain
\be
\label{eq:G3kernel}
G^{\textrm{\tiny (3)}}_{\textrm{\tiny mod}}(x,y;q,t)=
\frac{1}{p_V(q)}
\int_V
\mathrm{d}x_1\mathrm{d} x_2
\Sigma_t (x_1,y)\Sigma_t (x_2,y)
\int_{-\pi}^{\pi}\frac{d\alpha}{2\pi} 
e^{-\mathrm{i}\alpha q}
F^{\textrm{\tiny (3)}}(x,x_1,x_2;\alpha)\,,
\ee
where
\be
\label{eq:F3alpha}
F^{\textrm{\tiny (3)}}(x,x_1,x_2;\alpha)
\equiv
\langle\Omega|e^{\mathrm{i}\alpha \int_V j_+(x')
\mathrm{d}x'}
e^{\mathrm{i}\alpha \int_{V} j_-(y')
\mathrm{d}y'}
:\psi_+^\dagger(x_1)\psi_+(x_2):j_-(x)|\Omega\rangle
\,.
\ee
As mentioned in Sec.\,\ref{subsec:SRmdocorrfunc_FF} for the correlation function $F^{\textrm{\tiny (1)}}$, it is convenient to apply the bosonization rules (\ref{eq:bosonisationfermion}) and (\ref{eq:bosonisationcharge}) also in $F^{\textrm{\tiny (3)}}$. In this way, we obtain
\be
\label{eq:Falpha3_bosonised}
F^{\textrm{\tiny (3)}}(x,x_1,x_2;\alpha)
=\frac{
\left\langle\Omega\left|
\mathcal{V}^{\textrm{\tiny $(+)$}}_{\frac{\alpha}{2\pi}}(b_j)
\mathcal{V}^{\textrm{\tiny $(-)$}}_{\frac{\alpha}{2\pi}}(b_j)
\mathcal{V}^{\textrm{\tiny $(+)$}}_{-\frac{\alpha}{2\pi}}(a_j)
\mathcal{V}^{\textrm{\tiny $(-)$}}_{-\frac{\alpha}{2\pi}}(a_j) \colon \mathcal{V}^{\textrm{\tiny $(+)$}}_{1}(x_1) \mathcal{V}^{\textrm{\tiny $(+)$}}_{-1}(x_2) \colon
\partial \phi_-(x)
\right|\Omega\right\rangle}
{(2\pi)^2 \epsilon}\,,
\ee
where we have also used (\ref{eq:bosonisationFlux}).
Because of the factorization (\ref{eq:factorization}), we can compute separately the correlation functions involving the chiral fields $\phi_+$ and $\phi_-$ and then multiply them. Moreover, since in the case of our interest, both $\rho^{\textrm{\tiny $(+)$}}_V$ and $\rho_V^{\textrm{\tiny $(-)$}}$ are Gaussian, we can exploit the Wick's theorem to compute the two factors. Using also (\ref{eq:derivative-scalar}), (\ref{eq:vertexoperator}) and (\ref{eq:vertexoperatorantichiral}) and recalling the definition (\ref{eq:Lp_def}), we obtain
\be
\label{eq:F3alpha_v2}
F^{\textrm{\tiny (3)}}(x,x_1,x_2;\alpha)=
\frac{\alpha}{(2\pi)^3 }\frac{\bar{J}_p(x)}{x_1-x_2-\mathrm{i\epsilon}} e^{-\frac{\alpha^2}{2\pi^2}\ln L_p-\frac{\alpha}{2\pi}K_p(x_1,x_2)}
\,,
\ee
where $K_p(x_1,x_2)$ is defined in (\ref{eq:Kdef}) and
\be
\label{eq:Jbardef}
\bar{J}_p(x)\equiv \sum_{j=1}^p
\left(
\frac{b_j-a_j}{(b_j-x+\mathrm{i\epsilon})(x-a_j-\mathrm{i\epsilon})}
\right)
\,.
\ee
As a consistency check, we observe that the right-hand side of (\ref{eq:F3alpha_v2}) vanishes when $\alpha=0$, as can be also straightforwardly obtained from (\ref{eq:Falpha3_bosonised}) with $\alpha=0$ exploiting the factorization in the different chiral fields.
Computing the Fourier transform of (\ref{eq:F3alpha_v2}) within the saddle-point approximation around $\alpha=0$, we obtain
\bea
\label{eq:FTF3leadings}
&&\int_{-\pi}^{\pi}
\frac{d\alpha}{2\pi} 
e^{-\mathrm{i}\alpha q}
F^{\textrm{\tiny (3)}}(x,x_1,x_2;\alpha)
\\
&&=
\frac{p_V(q)}
{4\pi^2(x_1-x_2-\mathrm{i}\epsilon)}
\bigg[
\frac{-\bar{J}_p(x)}{2\ln L_p}
\bigg(
\mathrm{i}\pi q+\frac{K_p(x_1,x_2)}{2}\bigg)+
O\big[(\ln L_p)^{-2}\big]
\bigg]
\nonumber
\,,
\nonumber
\eea
where $p_V(q)$ is given in (\ref{eq:pq_Dirac}). Finally, plugging (\ref{eq:FTF3leadings}) back into (\ref{eq:G3kernel}), we find (\ref{eq:G3_Dirac_final}).
Notice that, differently from the expression (\ref{eq:F1alpha_v2}) of $F^{\textrm{\tiny (1)}}$, in (\ref{eq:F3alpha_v2}) there are no terms independent of $\alpha$ which multiply the exponential functions. This is what leads, after the Fourier transform, to the subleading behaviour in the UV cutoff of $G^{\textrm{\tiny (3)}}_{\textrm{\tiny mod}}$ compared to the one of $G^{\textrm{\tiny (1)}}_{\textrm{\tiny mod}}$.
A computation analogous to the one reported in this appendix can be performed for obtaining $G^{\textrm{\tiny (2)}}_{\textrm{\tiny mod}}(x,y;q,t)$ given in (\ref{G2_2dDirac}), leading to the result (\ref{eq:G2_Dirac_final}).

\end{appendix}

\bibliography{refsSRModularFlow}

\bibliographystyle{JHEP}

 \end{document}